\documentclass[a4paper,9pt]{extarticle}
\usepackage{hyperref}
\usepackage{fullpage}
\usepackage[utf8x]{inputenc}
\usepackage[titletoc]{appendix}
\usepackage{amssymb,amsmath}
\usepackage{floatrow}
\usepackage{stackengine}
\usepackage{graphicx}
\usepackage{wrapfig}
\usepackage{caption}
\usepackage{subcaption}
\usepackage{color}
\usepackage{multicol}
\usepackage{sidecap}
\usepackage{url}
\usepackage{color}
\usepackage{cite}
\usepackage{array}
\usepackage{tabulary}

\newcommand\numberthis{\addtocounter{equation}{1}\tag{\theequation}}

\newenvironment{Figure}
  {\par\medskip\noindent\minipage{\linewidth}}
  {\endminipage\par\medskip}
\newenvironment{Table}
  {\par\medskip\noindent\minipage{\linewidth}}
  {\endminipage\par\medskip}

\begin{document}

\title{\textbf{Inductively shunted transmon qubit with tunable transverse and longitudinal coupling}}
\author{Susanne Richer\textsuperscript{1}, Nataliya Maleeva\textsuperscript{2}, Sebastian T. Skacel\textsuperscript{2}, Ioan M. Pop\textsuperscript{2} and David DiVincenzo\textsuperscript{1}\\
\small{\textsuperscript{1}\textit{JARA Institute for Quantum Information, RWTH Aachen University, 52056 Aachen, Germany}}\\
\small{\textsuperscript{2}\textit{Physikalisches Institut, Karlsruhe Institute of Technology, 76131 Karlsruhe, Germany}}}
\maketitle

\renewcommand{\abstractname}{}
\begin{abstract}
We present the design of an inductively shunted transmon qubit with flux-tunable coupling to an embedded harmonic mode. This circuit construction offers the possibility to flux-choose between pure transverse and pure longitudinal coupling, that is coupling to the $\sigma_x$ or $\sigma_z$ degree of freedom of the qubit. While transverse coupling is the coupling type that is most commonly used for superconducting qubits, the inherently different longitudinal coupling has some remarkable advantages both for readout and for the scalability of a circuit. Being able to choose between both kinds of coupling in the same circuit provides the flexibility to use one for coupling to the next qubit and one for readout, or vice versa. We provide a detailed analysis of the system's behavior using realistic parameters, along with a proposal for the physical implementation of a prototype device.
\end{abstract}

\vspace{0.5cm}

\begin{multicols}{2}

\section{INTRODUCTION}

Superconducting qubits are among the most promising and versatile building blocks on the road to a functioning quantum computer. While qubit coherence times are getting better and better \cite{chang, barends, yan, Reagor2016, Minev2016}, it is still a challenge to couple qubits in a well-controlled manner, especially in circuit constructions that involve many qubits.
Most commonly, superconducting qubit architectures work with the so-called \textit{transverse coupling}, which involves coupling of the displacement degree of freedom of a resonator to the $\sigma_x$ degree of freedom of the qubit \cite{Wallraff2004, Blais2004}. While this well-studied coupling type is easy to implement and useful for dispersive readout, it is increasingly challenging to control in larger qubit architectures \cite{Riste2015, Barends2016, Paik2016, Versluis2017}. Unwanted cross-couplings degrade the circuit's performance, and the Purcell decay might limit the qubits' lifetimes \cite{Houck2008}.\\
In contrast to the usual transverse coupling, the inherently different \textit{longitudinal coupling} \cite{billangeon, didier, royer, Geller2015, Weber2017} means coupling to the $\sigma_z$ degree of freedom of the qubit. This coupling scheme can be used to implement strictly local interactions on a large-scale grid \cite{richer, billangeon} and plays an important role in the surface code architecture \cite{Fowler2012}. As shown in Ref.~\cite{didier}, longitudinal coupling might enable fast and efficient quantum nondemolition (QND) readout, while the usual dispersive readout is only approximately QND \cite{Boissonneault2009}. In addition, for a qubit coupled longitudinally to a resonator, the Purcell effect disappears as there is no dispersive shift. \\
In Ref.~\cite{richer} we presented a circuit design that consists of an inductively shunted transmon qubit with longitudinal coupling to an embedded harmonic mode.
We demonstrated that by applying a static external magnetic flux we can change the parity of the coupling between qubit and resonator mode in order to implement the desired longitudinal coupling. As will be shown here, the same architecture actually provides the possibility to flux-choose between pure longitudinal and pure transverse coupling, or have both at the same time. \\
While transverse coupling naturally appears in transmon-like circuit constructions, longitudinal coupling is usually much smaller and hardly ever the only coupling term present. The distinctive feature of the tunable design presented here is that the transverse coupling disappears when the longitudinal is maximal and vice versa. 
As opposed to other approaches, pure longitudinal coupling can be reached with moderate changes in the qubit frequency.
For conveniently chosen parameters, we show that longitudinal and transverse coupling have comparable values, while all other coupling terms can be suppressed.\\
In this paper we present a quantitative analysis of the flux dependence of all coupling terms for a realistic qubit-resonator circuit. 
We will also present an adapted alternative circuit, where coupling strength and anharmonicity scale better than in the original circuit and show how the anharmonicity and the coupling can be boosted by additional flux-biasing. \\
In Secs.~\ref{sec:quant} and \ref{sec:parameters}, we will have a closer look at the original circuit and explicitly derive the relevant quantities (frequencies, couplings and anharmonicities) as a function of the external flux. We will then turn our attention to the adapted circuit, which requires a numerical analysis (Sec.~\ref{sec:adaptation}). Using realistic parameters and obeying experimental constraints, we will provide a comparison of both circuits. Last but not least, in Sec.~\ref{sec:implementation} we present a proposal for an experimental device that will serve as a prototype for a first experiment. The sample, most of which can be fabricated using standard thin-film aluminum, could be embedded in a 3D wave\-guide with strong coupling to the resonator mode.

\section{INDUCTIVELY SHUNTED TRANSMON QUBIT}
\label{sec:quant}
Figure~\ref{fig:qubit_resonator} shows the circuit that implements an inductively shunted transmon qubit coupled to an embedded resonator, as introduced in Ref.~\cite{richer}.
The qubit essentially consists of a single Josephson junction with energy $E_{Jq}$, with a capacitance $C_q$ in parallel. We include the parallel plate capacitance of the qubit junction in $C_q$. The rest of the circuit is made up of two symmetric branches, each consisting of one or several Josephson junctions in parallel with a capacitance and an inductance. 
Similarly to the fluxonium qubit \cite{Manucharyan2009, Pop2014}, the inductive shunting protects the qubit from charge noise.
The qubit and resonator variables are chosen such that the superconducting phase differences across these two coupling branches are the sum and the difference of qubit and resonator variables, that is

\begin{align}
\varphi_q = \varphi_a - \varphi_b \qquad \varphi_r = \varphi_a + \varphi_b - 2\,\varphi_c,
\label{eq:variables}
\end{align}

where $\varphi_{abc}$ are the phases at the nodes of the circuit as depicted in Fig.~\ref{fig:qubit_resonator}.

\begin{Figure}
  \begin{center}
    \includegraphics[width=0.8\linewidth]{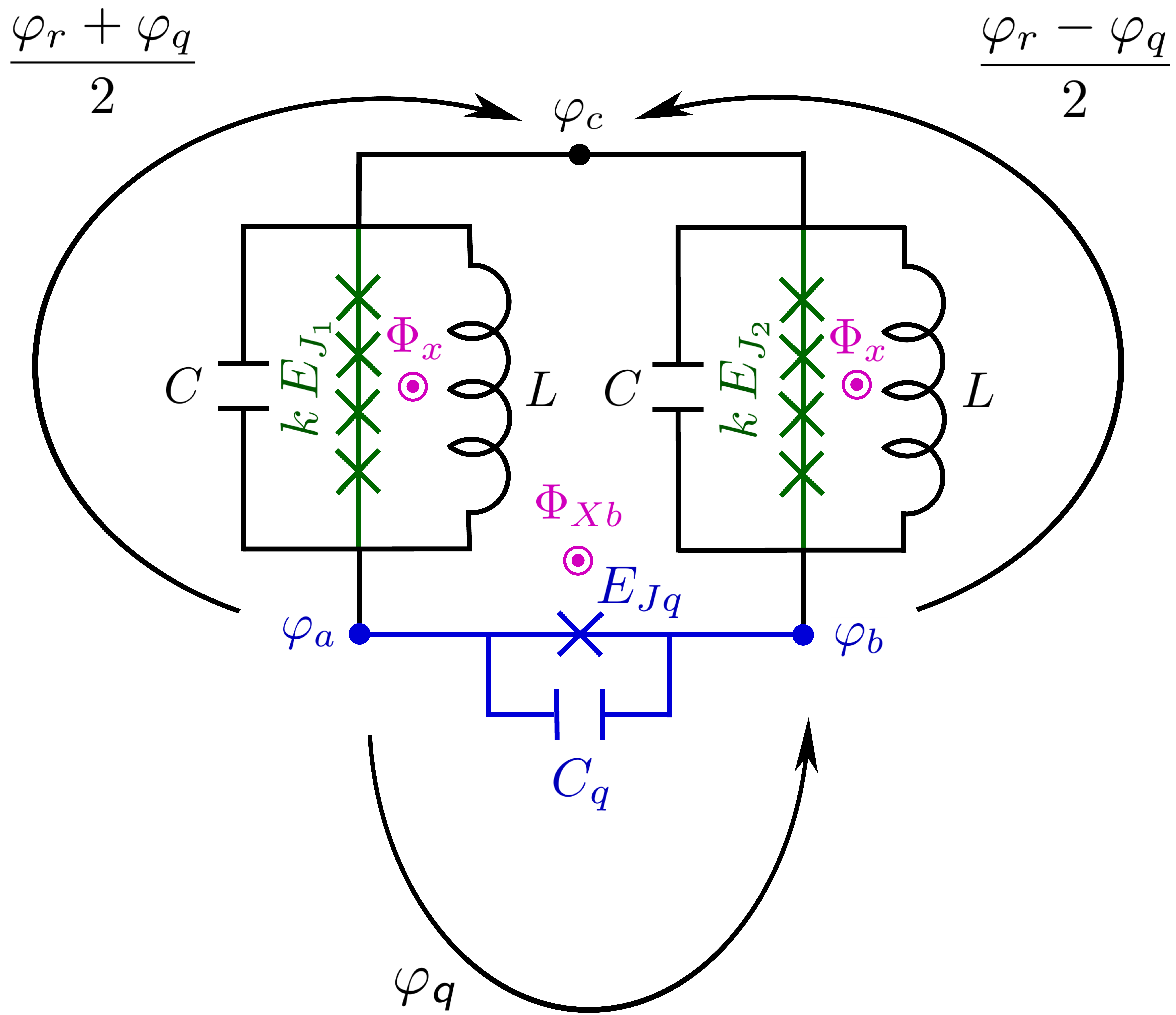}
    \captionof{figure}{Inductively shunted transmon qubit with the possibility to flux-choose between longitudinal and transverse coupling to an embedded resonator. The qubit mainly consists of a single Josephson junction (depicted in blue).
    }  
    \label{fig:qubit_resonator}
    \end{center}
\end{Figure}

Because of the left-right symmetry of the design, all coupling terms via the capacitances and inductances identically cancel out, and the coupling between qubit and resonator is only created by the coupling junctions (or junction arrays) $E_{J1}$ and $E_{J2}$. As shown in Ref.~\cite{richer}, the external flux $\Phi_x$ through the two coupling loops can be used to change the parity of the coupling term in order to implement longitudinal coupling. We will show here that the external flux $\Phi_x$ can also be used to tune between pure longitudinal and pure transverse coupling at conveniently chosen realistic parameters, and analyze the system as a function of this flux. Furthermore, we will allow for an additional external flux in the big loop $\Phi_{Xb}$ and show how the coupling and anharmonicity are boosted at $\Phi_{Xb} = \Phi_0/2$, where $\Phi_0$ is the magnetic flux quantum.
The kinetic energy of the qubit-resonator system is given by

\begin{align}
\mathcal T &= \left(\frac{\Phi_0}{2\pi}\right)^2 \left(\frac{2\,C_q+C}{4} \, \dot\varphi_q^2 + \frac{C}{4} \, \dot\varphi_r^2\right)
\label{eq:kin}
\end{align}

with the dimensionless phase variables as defined in Eq.~\ref{eq:variables}. Clearly, there is no coupling between qubit and resonator via the kinetic energy. The corresponding potential energy can be written as

\begin{align*}
\mathcal U &= \left(\frac{\Phi_0}{2\pi}\right)^2 \frac{1}{4L} \, (\varphi_q^2 + \varphi_r^2) - E_{Jq} \cos(\varphi_q + \varphi_{Xb})\\
& - k\,E_{J1} \cos\left(\frac{\varphi_r + \varphi_q}{2 k} + \frac{\varphi_x}{k}\right) \\
&- k\,E_{J2} \cos\left(\frac{\varphi_r - \varphi_q}{2 k} + \frac{\varphi_x}{k}\right),\numberthis
\label{eq:pot}
\end{align*}

where $\varphi_x = 2\pi \,\Phi_x/\Phi_0$ is the external flux through the two coupling loops and $\varphi_{Xb} =2\pi \,\Phi_{Xb}/\Phi_0$ is the external flux through the big loop, both rescaled to be dimensionless. We will assume $\varphi_{Xb} = 0$ in this section and consider its effect in Sec.~\ref{sec:fluxbias}. As depicted in Fig.~\ref{fig:qubit_resonator}, we might want to use arrays of $k$ equal Josephson junctions for the coupling branches in order to suppress the nonlinearity of the resonator as well as higher-order coupling terms. In Sec.~\ref{sec:array}, we will have a closer look at these arrays.  \\
The resonator is designed symmetrically, such that the coupling between qubit and resonator is only defined by the two coupling junction arrays $k\,E_{Ji}$, that is the last two lines in Eq.~\ref{eq:pot}. A trigonometric expansion leads to four different coupling terms, which we will classify by their parity. For the qubit, there are two terms with odd parity and two terms with even parity, meaning that these terms are an odd or even function in the qubit variable $\varphi_q$. The same is true for the resonator.\\
Longitudinal coupling involves the coupling of the displacement degree of freedom of the resonator to the $\sigma_z$ degree of freedom of the qubit. This means that the coupling term is an odd function in the resonator variable, and an even function in the qubit variable.
The longitudinal coupling $g_{zx}$ is maximal at $\varphi_x = k \, \pi/2$, when it is given by

\begin{align*}
 k\,E_{J\Sigma} \cos\left(\frac{\varphi_q}{2 k}\right)\sin\left(\frac{\varphi_r }{2 k}\right) \hat{=} \,\hbar \, g_{zx} \, \sigma_z (a^\dagger + a), \numberthis
\label{eq:coupz}
\end{align*}

which is an even function in the qubit variable and an odd function in the resonator variable. The longitudinal coupling term is proportional to the sum of the coupling junctions $E_{J\Sigma} = E_{J1} + E_{J2}$.\\
Transverse coupling on the other hand involves the coupling of the displacement degree of freedom of the resonator to the $\sigma_x$ degree of freedom of the qubit. This means that the coupling term is an odd function in both the resonator variable and the qubit variable. The transverse coupling $g_{xx}$ is maximal at zero flux, $\varphi_x = 0$, when it is given by

\begin{align*}
 k\,E_{J\Delta} \sin\left(\frac{\varphi_q}{2 k}\right)\sin\left(\frac{\varphi_r }{2 k}\right) \hat{=} \,\hbar \, g_{xx} \, \sigma_x (a^\dagger + a), \numberthis
\label{eq:coupx}
\end{align*}

which is an odd function both in the qubit and the resonator variable.
As opposed to the longitudinal coupling, the transverse coupling is proportional to the junction asymmetry $E_{J\Delta} = E_{J1} - E_{J2}$, which is designed to be about 3~-~8~\% of $E_{J\Sigma}$.
It is important to notice that the transverse term disappears at $\varphi_x = k \, \pi/2$, where the longitudinal coupling has its maximum, while the longitudinal coupling, on the other hand, disappears at zero flux. We will see later that for favorably chosen parameters, the other two coupling terms resulting from the expansion of Eq.~\ref{eq:pot}, $g_{xz}$ and $g_{zz}$, will be negligible, such that we can flux-choose between pure longitudinal and pure transverse coupling. 
Having both types of coupling in the same circuit gives us the flexibility to use one for coupling to the next qubit and one for readout, or vice versa. \\
In order to find expressions for the frequencies, anharmonicities, and couplings, we will employ the second quantization formalism. We go to the Hamiltonian representation and start by having a look at the quadratic terms of one variable, while the other is fixed at zero. (More accurately, it should be fixed at the potential minimum, which depends on the flux, as done in Sec.~\ref{sec:parameters}. We will see, however, that the formulas given here are a good approximation.) This treatment is similar to the one described in Ref.~\cite{nigg} about black-box quantization.\\
In order to quantize the qubit, we will at first treat it as a harmonic system and later on calculate its anharmonicity, that is its quartic deviation from a harmonic system.
A series expansion of the Hamiltonian around $\varphi_q = 0$ (at $\varphi_r = 0$) up to second order yields

\begin{align*}
\mathcal{H}_q = \frac{(2 e \, n_q)^2}{2\, C_q + C} &+ \frac{E_{Jq}}{2} \varphi_q^2 + \left(\frac{\Phi_0}{2\pi}\right)^2 \frac{1}{4L} \varphi_q^2 \\ 
&+ \frac{E_{J\Sigma}}{8k} \cos\left(\frac{\varphi_x}{k}\right) \, \varphi_q^2 \numberthis,
\label{eq:Hq1}
\end{align*}

where 

\begin{align}
n_q = \frac{1}{\hbar} \frac{\partial \mathcal{T}}{\partial \dot \varphi_q}
\label{eq:conj}
\end{align}

is a rescaled charge and the conjugate variable to $\varphi_q$. The potential energy of the qubit is governed by the Josephson energy of the qubit junction $E_{Jq}$, which is shunted by the inductance $L$.
In order to describe the flux dependence of Eq.~\ref{eq:Hq1}, we introduce the dimensionless coefficient 

\begin{align}
\eta =\frac{E_{J\Sigma}}{2k} \left(\frac{2\pi}{\Phi_0}\right)^2 L \cos\left(\frac{\varphi_x}{k}\right),
\label{eq:eta}
\end{align}

which disappears for pure longitudinal coupling, that is for $\varphi_x = k \, \pi/2$. This flux-dependence parameter is governed by the ratio between the Josephson energy of the coupling array and the energy of the inductance in parallel to it. 
Using Eq.~\ref{eq:eta} we define

\begin{align}
E_{Jq}^* = E_{Jq} +  \left(\frac{\Phi_0}{2\pi}\right)^2 \frac{1 + \eta}{2L}
\label{eq:EJeffective}
\end{align}

as the inductively shunted effective Josephson energy and 

\begin{align}
E_C = \frac{e^2}{2\,C_q + C}
\end{align}

as the charging energy of the qubit, and write 

\begin{align*}
\mathcal{H}_q &= 4\,E_C \, n_q^2 + \frac{E_{Jq}^*}{2} \varphi_q^2\numberthis.
\label{eq:Hq}
\end{align*}

We go to second quantization using

\begin{align*}
n_q &= \frac{1}{2}\sqrt[4]{\frac{E_{Jq}^*}{2 E_C}} i \, (c^\dagger - c)\\
\varphi_q &= \sqrt[4]{\frac{2 E_C}{E_{Jq}^*}} (c^\dagger + c) \numberthis
\label{eq:phiq}
\end{align*}

in Eq.~\ref{eq:Hq}, which yields 

\begin{align}
\mathcal{H}_q = \hbar \, \omega_q \, \left(c^\dagger c + \frac{1}{2}\right)
\end{align}

with the harmonic qubit frequency

\begin{align*}
\omega_q &= \frac{\sqrt{8 E_C E_{Jq}^*}}\hbar. \numberthis
\label{eq:delta}
\end{align*}

The quantization rules for the qubit given in Eq.~\ref{eq:phiq}, fulfill the commutation relation for the conjugate variables flux $\Phi_q$ and charge $Q_q$

\begin{align}
[\Phi_q, Q_q] = \left[\frac{\Phi_0}{2\pi}\varphi_q, 2e \,n_q\right] = \frac{i \, \hbar}{2} [c^\dagger + c, c^\dagger - c] = i \,\hbar
\end{align}

with $\Phi_0 = \pi \hbar/e$.
In order to determine whether the system can be treated as a qubit, we need to know its anharmonicity, 
that is its quartic deviation from a harmonic system. The fourth-order term in the potential energy (Eq.~\ref{eq:pot}) for $\varphi_r = 0$ is

\begin{align*}
&-\frac{1}{24} \left(E_{Jq} + \left(\frac{\Phi_0}{2\pi}\right)^2 \frac{\eta}{8\,k^2L}\right) \varphi_q^4\\
=&-\frac{E_C}{12 \, E_{Jq}^*} \left(E_{Jq} + \left(\frac{\Phi_0}{2\pi}\right)^2 \frac{\eta}{8\,k^2L}\right) (c^\dagger + c)^4\numberthis,
\end{align*}

using Eqs.~\ref{eq:eta} and \ref{eq:phiq}. Since

\begin{align}
\langle j | (a^\dagger + a)^4 | j \rangle = 6j^2 + 6j +3
\label{eq:quartic}
\end{align}

(see Ref.~\cite{koch}), where $|j \rangle$ are the Fock state eigenvectors, the energy of state $j$ up to fourth order is

\begin{align*}
E_j^{(q)} &= \sqrt{8E_C E_{Jq}^*} \left(j+\frac{1}{2}\right) \numberthis \label{eq:Ej}\\
&-\frac{E_C}{12 \, E_{Jq}^*} \left(E_{Jq} + \left(\frac{\Phi_0}{2\pi}\right)^2 \frac{\eta}{8\,k^2L}\right) (6j^2 + 6j +3).
\end{align*}

The quartic anharmonicity of the qubit is given by

\begin{align}
\alpha^{(q)} = \frac{E_{12}^{(q)} - E_{01}^{(q)}}\hbar = - E_C \, \frac{E_{Jq} + \left(\frac{\Phi_0}{2\pi}\right)^2 \frac{\eta}{8\, k^2L}}{\hbar E_{Jq}^*}
\label{eq:alpha}
\end{align}

with $E_{ij}^{(q)} = E_{j}^{(q)} - E_i^{(q)},$ which leads to a correction to the qubit frequency, that is

\begin{align}
\Delta = \frac{E_{01}^{(q)}}{\hbar} = \omega_q + \alpha^{(q)}.
\end{align}

We see that the qubit anharmonicity is governed by the charging energy of the qubit $E_C$ and the ratio between $E_{Jq}$ and $E_{Jq}^*$. Remembering again that $\eta = 0$ for pure longitudinal coupling at $\varphi_x = k \, \pi/2$, this is the same expression as the one given by Koch et al. in Ref.~\cite{koch} for the transmon anharmonicity, apart from the rescaling of the Josephson energy due to the inductive shunting (Eq.~\ref{eq:EJeffective}). In Ref.~\cite{koch} Koch et al. give an estimate for a minimal required relative anharmonicity of 

\begin{align}
\alpha_r^{(q)} = \frac{E_{12}^{(q)} - E_{01}^{(q)}}{E_{01}^{(q)}} \geq \frac{1}{200 \pi}.
\end{align}

As shown below, we can reach relative qubit anharmonicities which are one order of magnitude higher than this. As we will show in Sec.~\ref{sec:fluxbias}, the qubit anharmonicity can be significantly boosted using an additional flux bias $\varphi_{Xb}$ through the big loop (see Fig.~\ref{fig:qubit_resonator}).\\
In the two-level approximation, the qubit Hamiltonian is given by

\begin{align}
\mathcal{H}_q = \hbar \,\frac{\Delta}{2} \sigma_z.
\end{align}

For the resonator we follow the strategy used above and do a series expansion up to second order around $\varphi_r = 0$ (at $\varphi_q = 0$), that is

\begin{align*}
\mathcal{H}_r =\frac{(2 e \, n_r)^2}{C} +  \left(\frac{\Phi_0}{2\pi}\right)^2 \frac{1 + \eta}{4L}  \varphi_r^2  
=\frac{Q_r^2}{C} +   \frac{1 + \eta}{4L}  \Phi_r^2 
 \numberthis,
\label{eq:Hr}
\end{align*}

where the flux $\Phi_r$ and the charge $Q_r = 2e \,n_r$ are again conjugate variables that fulfill the commutation relation

\begin{align}
[\Phi_r, Q_r] = \left[\frac{\Phi_0}{2\pi}\varphi_r, 2e \,n_r\right] = \frac{i \, \hbar}{2} [a^\dagger + a, a^\dagger - a] = i \,\hbar.
\end{align}

The quantization step is done by inserting

\begin{align*}
Q_r &=  2e \, n_r = \sqrt{\frac{\hbar}{2  Z_{0}}} i \, (a^\dagger - a)  \\
\Phi_r &= \frac{\Phi_0}{2\pi}\varphi_r= \sqrt{\frac{\hbar  Z_{0}}{2}} (a^\dagger + a)\numberthis
\label{eq:phir}
\end{align*}

(see Ref.~\cite{devoret}) in Eq.~\ref{eq:Hr} and choosing the characteristic impedance $Z_{0}$ such that the Hamiltonian has the form

\begin{align}
\mathcal{H}_r = \hbar \, \omega_r \, \left(a^\dagger a + \frac{1}{2}\right),
\end{align}

i.e. such that the non-diagonal terms disappear. This is satisfied for

\begin{align}
Z_{0} = 2\sqrt{\frac{L}{C(1+\eta)}},
\end{align}

which directly gives

\begin{align}
\omega_r = \sqrt{\frac{1 + \eta}{L \,C}}
\label{eq:omegar}
\end{align} 

for the resonator frequency. We see that $\omega_r$ acquires a flux-dependence due to $\eta$ (see Eq. \ref{eq:eta}). However, the effect of $\eta$ can be suppressed by increasing the number of junctions $k$ in the array (see Sec.~\ref{sec:array}).\\ 
In order to verify whether our resonator can really be treated as a harmonic system, we will calculate its anharmonicity. Using Eq.~\ref{eq:phir}, the fourth-order term in the potential energy (Eq.~\ref{eq:pot}), again for $\varphi_q = 0$, can be written as

\begin{align*}
&- \left(\frac{\Phi_0}{2\pi}\right)^2 \frac{\eta}{192\, k^2 L} \varphi_r^4\\
=& - \left(\frac{2\pi}{\Phi_0}\right)^2 \frac{\hbar^2\eta}{192\, k^2 C(1 + \eta)} (a^\dagger + a)^4 \numberthis.
\end{align*}

Using again Eq.~\ref{eq:quartic}, the energy of state $j$ up to fourth order is

\begin{align*}
E_j^{(r)} = \hbar \, \omega_r \left(j+\frac{1}{2}\right) - \left(\frac{2\pi}{\Phi_0}\right)^2 \frac{\hbar^2\eta \, (6j^2 + 6j +3)}{192\, k^2 C(1 + \eta)}\numberthis. 
\end{align*}

The anharmonicity of the resonator is then given by

\begin{align*}
\alpha^{(r)} &= \frac{E_{12}^{(r)} - E_{01}^{(r)}}\hbar = \frac{\eta \,\pi^2 \hbar }{\eta \,\pi^2 \hbar  - 4 k^2 (1+\eta)^\frac{3}{2} \sqrt{C/L}} \numberthis,
\label{eq:alphares}
\end{align*}

where again $E_{ij}^{(r)} = E_{j}^{(r)} - E_i^{(r)}$. We see that the anharmonicity of the resonator is proportional to the parameter $\eta$. Remarkably, $\eta$ is zero at the longitudinal coupling point $\varphi_x = k \, \pi/2$, where the resonator anharmonicity goes through zero and changes its sign. This remains true when we include higher order terms, as all series terms in the potential energy in $\varphi_r$ from third order onward are proportional to $\eta$. If we want to work with a static system with pure longitudinal coupling, we can thus assume our resonator to be perfectly harmonic. At any other point in flux, though, we should choose our parameters carefully, in order to ensure that $\alpha^{(r)}$ remains small.\\
Now, we would like to have a look at the coupling terms, taking into account the four terms with different parities, as mentioned above. We will do a series approximation of the potential energy up to second order in both $\varphi_r$ and $\varphi_q$ around zero, which is assumed to be the potential minimum (see Sec.~\ref{sec:parameters} for a more exact numerical treatment).
For two identical coupling junctions (or coupling arrays), only $\sigma_z$-type coupling terms are possible, as all uneven terms in $\varphi_q$ cancel out. This means that the $\sigma_z$ coupling terms are proportional to the sum of the coupling junctions $E_{J\Sigma}$, while the $\sigma_x$ coupling terms are proportional to their difference $E_{J\Delta}$. \\
At zero flux $\varphi_x = 0$, we find the transverse coupling, which we call $g_{xx}$ as it has odd parity both in $\varphi_q$ and $\varphi_r$. It is

\begin{align*}
\frac{E_{J\Delta}}{4\,k}   \, \varphi_q \, \varphi_r \cos\left(\frac{\varphi_x}{k}\right) \hat{=} \, \hbar \, g_{xx} \, \sigma_x (a^\dagger + a) \numberthis
\end{align*}

with

\begin{align}
g_{xx} =\frac{E_{J\Delta}}{2\,k \sqrt{\hbar}}   \sqrt[4]{\frac{2 E_C}{E_{Jq}^*}}  \frac{\pi}{\Phi_0} \sqrt[4]{\frac{L}{C}\frac{1}{1 + \eta}} \cos\left(\frac{\varphi_x}{k}\right).
\label{eq:gxx}
\end{align}

There is a competing term with a similar flux dependence, which has even parity in both $\varphi_q$ and $\varphi_r$, namely

\begin{align*}
-\frac{E_{J\Sigma}}{64\,k^3}  \, \varphi_q^2 \, \varphi_r^2 \cos\left(\frac{\varphi_x}{k}\right) \hat{=} \, \hbar \,  g_{zz} \, \sigma_z \, (a^\dagger + a)^2 \numberthis
\end{align*}

with

\begin{align}
g_{zz} &=-\frac{E_{J\Sigma}}{16\,k^3}   \sqrt{\frac{2 E_C}{ E_{Jq}^*}}  \left(\frac{\pi}{\Phi_0}\right)^2 \sqrt{\frac{L}{C}\frac{1}{1 + \eta}} \cos\left(\frac{\varphi_x}{k}\right).
\label{eq:gzz}
\end{align}

At $\varphi_x = k \, \pi/2$ both $g_{xx}$ and $g_{zz}$ vanish, while two other coupling terms are at their joint maximum. One is the longitudinal coupling term

\begin{align*}
-\frac{E_{J\Sigma}}{16\,k^2}   \, \varphi_q^2 \, \varphi_r \sin\left(\frac{\varphi_x}{k}\right) \hat{=}\, \hbar \, g_{zx} \, \sigma_z (a^\dagger + a) \numberthis
\end{align*}

with

\begin{align}
g_{zx} =-\frac{E_{J\Sigma}}{8\,k^2 \sqrt{\hbar}}   \sqrt{\frac{2 E_C}{E_{Jq}^*}}  \frac{\pi}{\Phi_0} \sqrt[4]{\frac{L}{C}\frac{1}{1 + \eta}} \sin\left(\frac{\varphi_x}{k}\right),
\label{eq:gzx}
\end{align}

where the $z$ in $g_{zx}$ stands for even qubit parity and the $x$ for odd resonator parity.
The competing $\sigma_x$ term has opposite parity

\begin{align*}
-\frac{E_{J\Delta}}{16\,k^2}  \, \varphi_q \, \varphi_r^2 \sin\left(\frac{\varphi_x}{k}\right) \hat{=} \, \hbar \, g_{xz} \, \sigma_x \, (a^\dagger + a)^2 \numberthis
\end{align*}

with

\begin{align}
g_{xz} &=-\frac{E_{J\Delta}}{4\,k^2}   \sqrt[4]{\frac{2 E_C}{ E_{Jq}^*}}  \left(\frac{\pi}{\Phi_0}\right)^2 \sqrt{\frac{L}{C}\frac{1}{1 + \eta}} \sin\left(\frac{\varphi_x}{k}\right).
\label{eq:gxz}
\end{align}

We will see later that for conveniently chosen parameters, only the transverse coupling $g_{xx}$ (which is the lowest order term) and the longitudinal coupling $g_{zx}$ play a role. The $g_{zz}$ term is the highest order term and therefore much smaller than the others. The $g_{xz}$ term is of the same order as the longitudinal coupling, but is suppressed by the small junction asymmetry $E_{J\Delta} \ll E_{J\Sigma}$. As the flux dependences of longitudinal and transverse coupling have a quadrature relation to one another, each is at its maximum when the other disappears and vice versa. \\
In Ref.~\cite{didier}, similar expressions are obtained for transverse and longitudinal coupling between a qubit and a resonator. 
However, the qubit considered there is a split transmon with a single flux loop, in which the two Josephson junctions play the role of qubit junctions and coupling junctions at the same time. 
For small junction asymmetries ($d = 0.02$ in Ref.~\cite{didier}), switching to pure longitudinal coupling would be accompanied by a much larger change in the qubit frequency compared to our design, where the roles of the qubit and coupling junctions are separated.\\
As recently shown by Hutchings et al. in Ref.~\cite{Hutchings2017}, the qubit dephasing rate of transmons is proportional to the sensitivity of the qubit frequency to the external flux.
Staying in a regime with moderate flux tunability in the range of hundreds of MHz (see Secs.~\ref{sec:parameters} and \ref{sec:adaptation}), our qubit should be nearly unaffected by flux noise. Note that this is true independent of the coupling junction asymmetry. 
It is important that in our design the coupling can be tuned by applying an external flux through the two smaller coupling loops, while an additional flux through the big qubit loop can be used to boost the anharmonicity. For the bigger loop we consider the two cases $\Phi_{Xb} = 0$ and $\Phi_{Xb} = \Phi_0/2$, both of which are sweet-spots with respect to flux noise.

\subsection{EFFECT OF ASYMMETRIES}
The capacitances and inductances in the design shown in Fig.~\ref{fig:qubit_resonator} are supposed to be symmetric, such that the coupling between qubit and resonator is only created by the coupling junctions $E_{J1}$ and $E_{J2}$. This has the advantage that the resulting transverse coupling (Eq.~\ref{eq:gxx}) is flux-dependent and goes through zero at $\varphi_x = k \, \pi/2$, which leads to pure longitudinal coupling. Transverse coupling terms caused by asymmetric inductances or capacitances would, however, be independent of the external flux. While the capacitances in our design (see Sec.~\ref{sec:implementation}) can be fabricated very accurately, the asymmetry in the inductances could be in the neighborhood of $\delta L = (L_1 - L_2)/(L_1 + L_2) \sim 0.01$. To first order in $\delta L$, the frequencies and anharmonicities of qubit and resonator are unaffected by this asymmetry. The same is true for  the coupling terms with even qubit parity, such as the longitudinal coupling $g_{zx}$. The coupling terms with odd qubit parity, such as the transverse coupling $g_{xx}$, contain, however, a term proportional to $\delta L$. To first order in $\delta L$, the total transverse coupling is given by

\begin{align}
g_{xx}^\text{total} = g_{xx}^\text{asym} + |g_{xx}^\text{sym}| \cos\left(\frac{\varphi_x}k\right),
\label{eq:coup_asym}
\end{align}

where the symmetric flux-dependent part is given by Eq.~\ref{eq:gxx}.
The asymmetric part due to the unequal inductances is a constant offset independent of flux. From Eq.~\ref{eq:coup_asym} it is clear that as long as $|g_{xx}^\text{sym}| / |g_{xx}^\text{asym}| > 1$, there is still a point in flux where the total transverse coupling $g_{xx}^\text{total}$ goes through zero. For a significant $\delta L$ of a few percent, this point might be considerably shifted from the ideal value of $\varphi_x = k \, \pi/2$, where the longitudinal coupling is maximal. However, this shift can be suppressed by increasing the coupling junction asymmetry $d$, as $|g_{xx}^\text{sym}| / |g_{xx}^\text{asym}|$ is proportional to $d/\delta L$ \cite{mythesis}.

\section{CHOICE OF PARAMETERS}
\label{sec:parameters}
One approximation we made in deriving the formulas above was to assume that the potential energy minimum is always at $\varphi_q = \varphi_r = 0$. This is crucial as all important quantities were derived using series approximations around the potential minimum. However, looking closely at the potential function given in Eq.~\ref{eq:pot}, we see that this is only true at fluxes $\varphi_x = \mu \, k \, \pi$ for integer multiples $\mu$, but nowhere in between. The exact position of the minimum depends of course strongly on the chosen parameters. The solution is thus to numerically determine the potential energy minimum for a given set of parameters (including the external fluxes) and calculate the frequencies, anharmonicities, and couplings again by series approximations around this potential minimum. As we will see in the next section, the formulas given above are a good approximation. Though they cannot capture the flux dependence exactly, they always give the right values at fluxes $\varphi_x = \mu \, k \, \pi$.
This numerical treatment becomes especially important, when we allow for a flux in the big loop $\varphi_{Xb}$, which can be used to boost the anharmonicity (see Sec.~\ref{sec:fluxbias}).\\
A possible experiment to verify the model could be a measurement of the qubit-resonator dispersive shift as a function of the external flux through the coupling loops $\varphi_x$. While the transverse coupling $g_{xx}$ leads to a qubit state dependence of the resonator frequency, the longitudinal coupling $g_{zx}$ does not. This means that the qubit-resonator dispersive shift should disappear at $\varphi_x = k \, \pi/2$, where $g_{xx}$ goes through zero and we have pure longitudinal coupling.\\
In this section, we will discuss how to favorably choose the parameters for such an experiment given the constraints of the real system, keeping in mind that all important quantities are flux-dependent. For example, we will require the resonator frequency to always stay in the range of $\omega_r/(2\pi) = 6 - 8$ GHz, which is a convenient microwave range, recently used in the setup of Ref.~\cite{Kou2017} to perform multiplexed quantum readout. We want the qubit frequency to be well separated from the resonator frequency, as any overlap could lead to unwanted cross talk. 
As mentioned above, it is necessary to stay in a regime with moderate flux tunability of the qubit frequency in order to avoid dephasing due to flux noise.
These being hard constraints, our goal is that the qubit anharmonicity should be as high as possible, while the anharmonicity of the resonator should be negligible.
As we are aiming here for a system where we can flux-choose between transverse and longitudinal coupling, we will choose our parameters such that the longitudinal coupling is as high as possible, while the transverse coupling should be comparable. Note that the transverse coupling can be easily controlled via the junction asymmetry. All other coupling terms should be negligible in order to have pure longitudinal or transverse coupling.\\
We will treat the cases of single coupling junctions and coupling junction arrays separately as they require different restrictions on the parameters. In addition, for both cases we will examine the effect of a flux-biasing of $\varphi_{Xb} = \pi$ in the big loop.

\subsection{CASE ONE: SINGLE COUPLING JUNCTIONS}
\label{sec:single}
In the following, we will treat the case of single coupling junctions, which means we set $k = 1$ in all the formulas from Sec.~\ref{sec:quant}.
Looking at the expression for the resonator frequency given in Eq.~\ref{eq:omegar}, it becomes clear that the absolute value of $\eta$ should not become bigger than $1$, since within our series approximation $\omega_r$ would not be well-defined. Following from Eq.~\ref{eq:eta}, it is clear that $\omega_r$ will have a maximum at zero flux and a minimum at $\varphi_x = \pi$. In order to ensure that it always stays between 6 and 8 GHz, we can fix the capacitance $C$ in terms of $L$ and $|\eta|$, such that $\omega_r/(2\pi) = 8$ GHz at its maximum, and then define a maximal inductance $L_\text{max}$, such that $\omega_r/(2\pi) \geq 6$ GHz at its minimum. With these hard constraints and the goal of having high coupling and qubit anharmonicity, while keeping the qubit frequency well separated from the resonator frequency, we tried out different parameter values until we found the \textit{optimal} solution. The junction asymmetry $d = E_{J\Delta}/E_{J\Sigma}$ is chosen such that the maximal transverse coupling $g_{xx}$ is approximately as big as the maximal longitudinal coupling $g_{zx}$.

\begin{Table}
\begin{center}
    \begin{tabular}{| >{\centering}m{.16\linewidth} | c || c | c |}  
    \hline
    \multicolumn{2}{|c||}{Parameters} & \multicolumn{2}{|c|}{Results}\\
    \hline
    $E_{Jq}$ & $h$ 10 GHz & $\omega_r/(2\pi)$ & 6.2 - 8 GHz \\ \hline
    $E_{J\Sigma}$ & $h$ 20 GHz & $\Delta/(2\pi)$ & 5.4 - 6.4 GHz\\ \hline
     $E_{J\Delta}/E_{J\Sigma}$ & 0.08 & $g_{zx}^\text{max}/(2\pi)$ & 53 MHz \\ \hline
     $C$ & 114 fF & $g_{xx}^\text{max}/(2\pi)$ & 49 MHz \\ \hline
     $C_q$ & 70 fF & $g_{zz}^\text{max}/(2\pi)$ & 5 MHz \\ \hline
   $L$ & 4.5 nH & $g_{xz}^\text{max}/(2\pi)$ &  6 MHz \\ \hline
   $L_\text{max}$ & 4.9 nH & $|\alpha_r^{(q)}|$ & 0.8 - 1.1\% \\ \hline
   $L_\text{crit}$ & 5.6 nH & $|\alpha_r^{(r)}|$ &  $\leq$ 0.5\% \\ \hline
    \end{tabular}
    \captionof{table}{A good choice of parameters for the single coupling junction case ($k=1$) at zero flux through the big loop $\varphi_{Xb} = 0$. $L$ needs to be less than or equal to $L_\text{max}$ to ensure that the resonator frequency stays in the 6 - 8 GHz range and less than $L_\text{crit}$ in order to avoid a double-well potential for all possible values of flux (compare Sec.~\ref{sec:fluxbias}).
    On the right we show the frequencies, anharmonicities, and couplings, which vary with the flux in the coupling loops.}
    \label{tb:k1}
\end{center}
\end{Table}

Table~\ref{tb:k1} shows the chosen parameters and the frequencies, anharmonicities, and couplings they lead to. Figure~\ref{fig:freqk1} shows the frequencies of qubit and resonator as a function of the flux through the coupling loops. As we can see, they always stay well separated. 

\begin{Figure}
  \begin{center}
    \includegraphics[width=\linewidth]{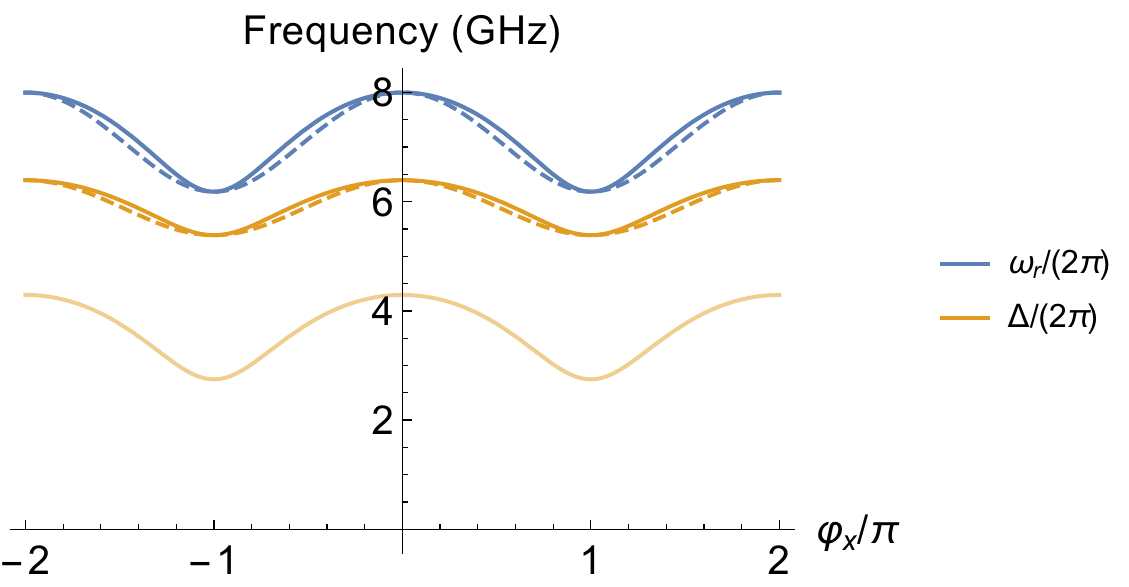}
    \captionof{figure}{The frequencies of qubit and resonator as a function of the (reduced) flux through the coupling loops $\varphi_x$. Solid lines show accurate numerical results, dashed lines show the predictions using the formulas from Sec.~\ref{sec:quant}, the lighter color curve shows results at a flux $\varphi_{Xb} = \pi$ in the big loop. While the resonator frequency is not affected by the large-loop flux-biasing, the qubit frequency experiences a drop.}  
    \label{fig:freqk1}
    \end{center}
\end{Figure}

\begin{Figure}
  \begin{center}
    \includegraphics[width=\linewidth]{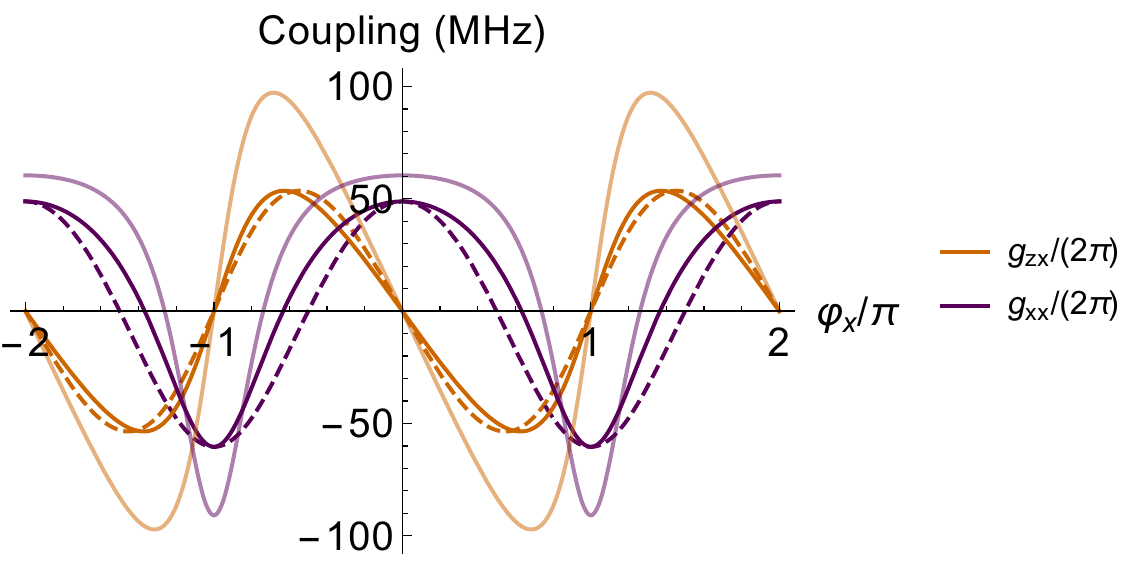}
    \captionof{subfigure}{Longitudinal ($g_{zx}$) and transverse coupling ($g_{xx}$).}  
    \label{fig:coupk1_1}
    \includegraphics[width=\linewidth]{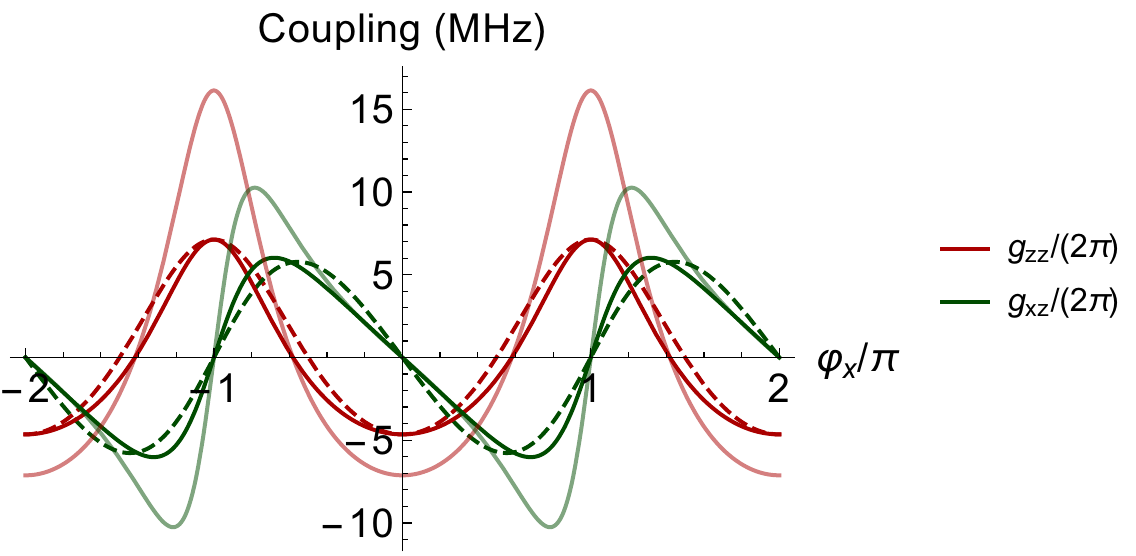}
    \captionof{subfigure}{Unwanted coupling terms, note the change of scale.}  
    \label{fig:coupk1_2}
    \captionof{figure}{The four most important coupling terms as a function of the (reduced) flux through the coupling loops $\varphi_x$. Solid lines show accurate numerical results, dashed lines show the predictions using the formulas from Sec.~\ref{sec:quant}, the lighter color curves show results at a flux $\varphi_{Xb} = \pi$ in the big loop. The longitudinal coupling is almost doubled due to the large-loop flux-biasing, while the point where the transverse coupling disappears is considerably shifted. The longitudinal coupling always disappears exactly at multiples of $\varphi_x = k\, \pi$.}
    \label{fig:coupk1}
    \end{center}
\end{Figure}

Solid lines in the figure show our accurate numerical results including the effect of the flux-dependent potential energy minimum (see above), dashed lines show the predictions using the formulas from Sec.~\ref{sec:quant}. While the predictions are always accurate at the maxima and minima, they deviate slightly in between. The lighter color curve shows results at a flux $\varphi_{Xb} = \pi$ through the big loop (see Sec.~\ref{sec:fluxbias}).
\\
While both the longitudinal and the transverse coupling reach values above 50 MHz, the spurious terms are far from being negligible (see Fig.~\ref{fig:coupk1}). The unwanted $g_{xz}$ coupling reaches 11~\% of the longitudinal coupling $g_{zx}$ at their joint maximum, while the unwanted $g_{zz}$ reaches 9~\% of the transverse coupling $g_{xx}$ at their maximum. The dashed lines show again the predictions using the formulas from Sec.~\ref{sec:quant}. We see that they are a good but not perfect approximation. In particular, the point where the transverse coupling disappears and the longitudinal coupling peaks is shifted. The lighter color curves show what happens due to the large-loop flux-biasing (see Sec.~\ref{sec:fluxbias}).

\begin{Figure}
  \begin{center}
    \includegraphics[width=\linewidth]{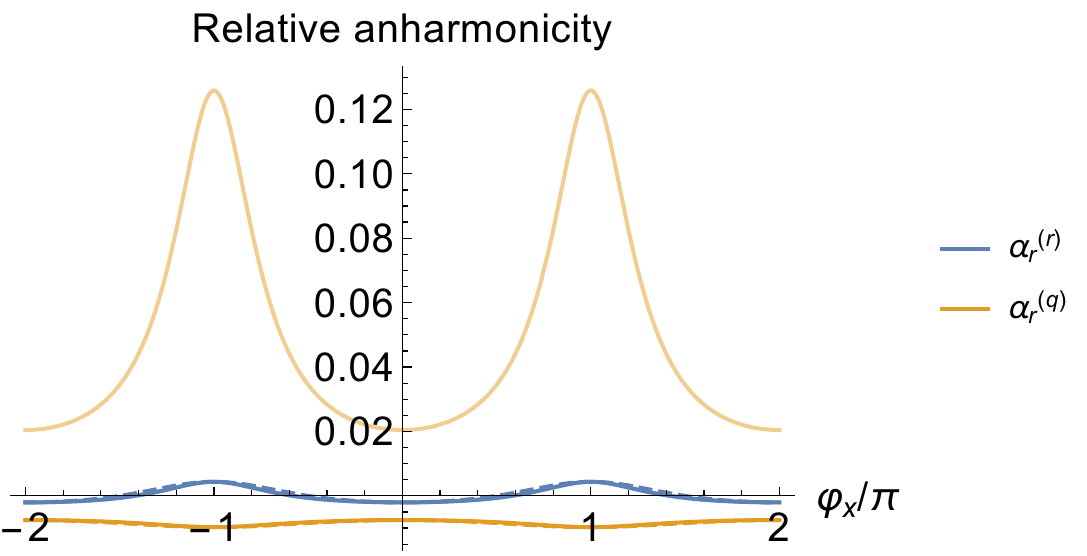}
    \captionof{figure}{The relative anharmonicities of qubit and resonator
     as a function of the (reduced) flux through the coupling loops $\varphi_x$. Solid lines show accurate numerical results, dashed lines show the predictions using the formulas from Sec.~\ref{sec:quant}. Note that in this plot the predictions are indistinguishable from the numerical results. The lighter color curve shows results at a flux $\varphi_{Xb} = \pi$ in the big loop.}  
    \label{fig:anhk1}
    \end{center}
\end{Figure}

The resonator anharmonicity is clearly a problem, as it is almost as large as the qubit anharmonicity (see Fig.~\ref{fig:anhk1}). While it goes through zero almost exactly when the transverse coupling disappears, it is much too high at all other values of flux. The lighter color curve shows again results at a flux $\varphi_{Xb} = \pi$ in the big loop, to be discussed now. 

\subsection{FLUX-BIASING}
\label{sec:fluxbias}
Applying a flux of $\varphi_{Xb} = \pi$ through the big loop as suggested in Eq.~\ref{eq:pot} has a very interesting effect on the circuit's behavior. It changes the qubit spectrum, but does not affect the resonator. As shown in Fig.~\ref{fig:freqk1}, the qubit frequency drops to 2.5 - 4~GHz, while the resonator frequency remains unchanged. The qubit anharmonicity is now positive and boosted up to 17~\% (see Fig.~\ref{fig:anhk1}). The coupling also gets a boost and is approximately doubled (see Fig.~\ref{fig:coupk1}).\\
Now, what exactly happens, when we put a flux through the big loop? Looking again at the expression for the qubit frequency (Eq.~\ref{eq:delta}) and its derivation, it becomes clear that a flux $\varphi_{Xb} = \pi$ through the big loop corresponds to the transition $E_{Jq} \to - E_{Jq}$ in the potential function. The qubit potential thus consists primarily of a parabola with its minimum at $\varphi_q = 0$ (due to the inductive part) and a cosine with a maximum at $\varphi_q = 0$ (due to the qubit function). Clearly, this could lead to a double-well potential, similar to flux qubits \cite{yan, Chiorescu2003}. This is a case which we want to avoid - we have chosen not to explore flux (i.e. persistent-current) qubits, and thus all our analysis is designed for a single well treatment. Looking at the expression for the qubit frequency with $E_{Jq} \to - E_{Jq}$, we find

\begin{align}
\Delta^\pi = \frac{\sqrt{8 E_C (E_L(1 + \eta)  - E_{Jq})}}{\hbar},
\end{align}

where $E_L = (\Phi_0/(2\pi))^2/(2L)$ is the energy of the inductance. This explains the drop in the qubit frequency shown in Fig.~\ref{fig:freqk1}. It also shows that we need to make sure that $E_L (1+\eta)$ is always bigger than $E_{Jq}$, such that the qubit frequency remains well-defined and we avoid the double-well potential. This implies another critical (maximal) inductance, which is

\begin{align}
L_\text{crit} = \left(\frac{\Phi_0}{2\pi}\right)^2 \frac{1 + \eta}{2 E_{Jq}}.
\label{eq:lcrit}
\end{align}

For the parameters used here, $L_\text{crit}$ is, however, bigger than the $L_\text{max}$ defined in Sec.~\ref{sec:single}, such that it does not limit the permitted parameter space any further (see Tab.~\ref{tb:k1}). The qubit anharmonicity also changes considerably due to the flux in the big loop. It is

\begin{align}
\alpha^{(q) \pi} = - E_C \frac{E_L \frac{\eta}{4k^2} - E_{Jq} }{\hbar (E_L (1 + \eta) - E_{Jq})}.
\label{eq:anhpi}
\end{align}

With the parameters from Tab.~\ref{tb:k1}, $E_{Jq}$ is approximately half of $E_L$, which means that the denominator in Eq.~\ref{eq:anhpi} is positive (this is actually required by Eq.~\ref{eq:lcrit}), while the numerator is negative, yielding a positive anharmonicity. Figure~\ref{fig:anh3dk1} shows that the qubit anharmonicity has a steep maximum at $\varphi_{Xb} = \varphi_x = \pi$. While it is negative in a large range around $\varphi_{Xb} = 0$, it changes sign when approaching $\varphi_{Xb} = \pi$. This implies that there is a point in between where the qubit anharmonicity is zero. The qubit with $\varphi_{Xb}=\pi$ is reminiscent of the capacitively shunted flux qubit (CSFQ) (see Ref.~\cite{yan}).\\
While both the coupling and the qubit anharmonicity are quite strong at flux $\pi$ through the big loop, the non-negligible unwanted coupling terms $g_{xz}$ and $g_{zz}$ remain a problem, as well as the high nonlinearity of the resonator. Plus, we would like to have a system that also performs at zero flux.
We will therefore try out different adaptations, one simply being arrays of Josephson junctions instead of the single coupling junctions.

\begin{Figure}
  \begin{center}
    \includegraphics[width=\linewidth]{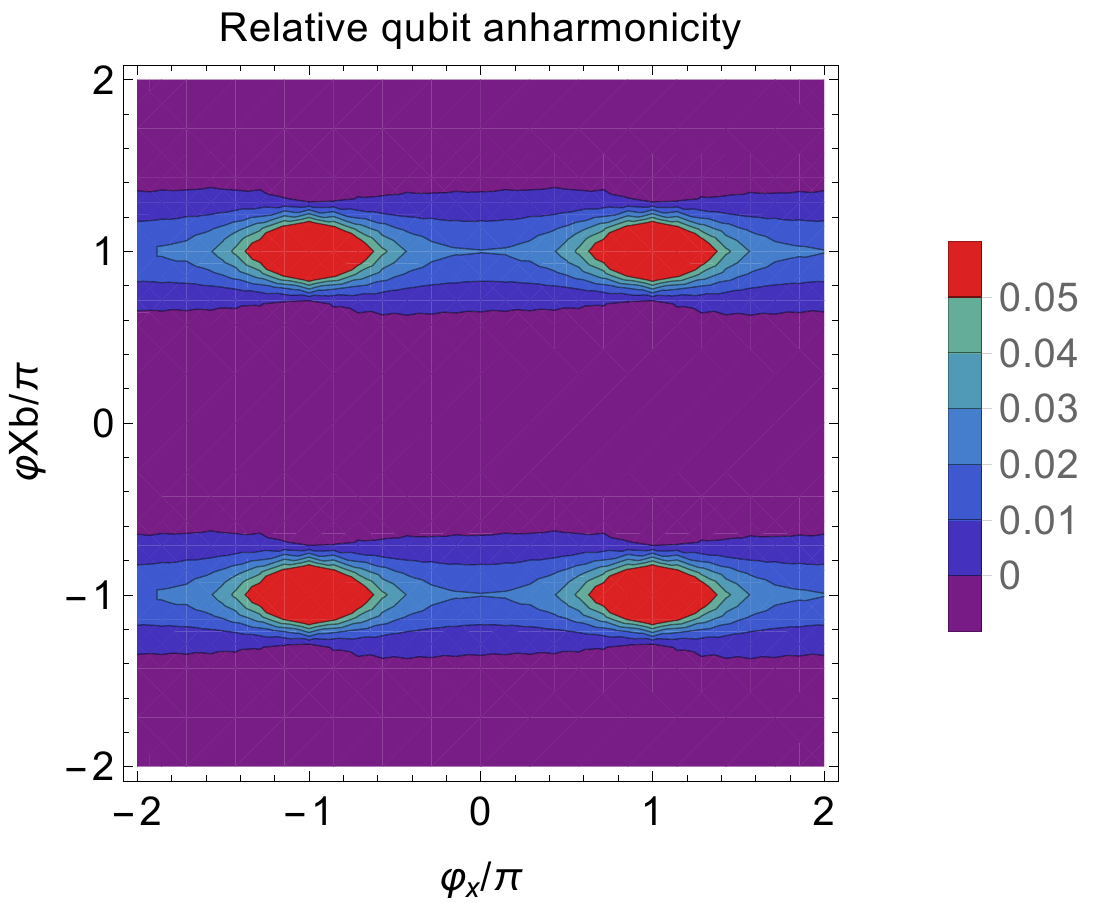}
    \captionof{figure}{Relative anharmonicity of the qubit as a function of the (reduced) fluxes through the coupling loops $\varphi_x$ and through the big loop $\varphi_{Xb}$. While the anharmonicity is negative in a large range around $\varphi_{Xb} = 0$ (dark blue region), it changes sign when approaching $\varphi_{Xb} = \pi$ and has a steep maximum at $\varphi_{Xb} = \varphi_x = \pi$.}  
    \label{fig:anh3dk1}
    \end{center}
\end{Figure}

\subsection{CASE TWO: COUPLING JUNCTION ARRAYS}
\label{sec:array}
When we want to substitute the single coupling junctions by coupling arrays, there are a few things we have to take into account.  Assuming that all junctions in such an array are equal, we can describe the potential energy of an array of $k$ junctions as

\begin{align}
\mathcal{U}_k = - \sum_{i=1}^k E_{Ji} \cos(\varphi_i) = - k E_J \cos\left(\frac{\varphi + m \,2\pi}{k}\right),
\label{eq:u_array}
\end{align}

where $\varphi = \sum_i \varphi_i$ is the total phase over the junction array and $m \in \mathbb{Z}$ is the integer number of flux quanta in a loop formed by the junctions, thereby numbering the metastable solutions for $\varphi$. For the coupling scheme to work, we require $m$ to be constant in time over long durations.
As described in Ref.~\cite{masluk, Matveev2002}, so-called phase-slip events, that is integer changes in $m$, can be detected by jumps in the frequency of the system. However, phase slips are suppressed by choosing a large $E_{Ji}/E_{Ci}$ ratio for each individual junction and time spans on the order of hours or days with constant $m$ can be realistically achieved \cite{Pop2010}. We will take $m$ to be zero here and expand for large $k$, finding

\begin{align}
\mathcal{U}_k \approx - k E_J + \frac{E_J}{2 k} \varphi^2  = - k E_J + \left(\frac{\Phi_0}{2\pi}\right)^2 \frac{1}{2L_J} \, \varphi^2
\end{align}
 
with an effective inductance of $L_J = k\, (\Phi_0/(2\pi))^2 /E_J$. The effective inductance of such an array is thus proportional to the number of junctions $k$.

\begin{Figure}
  \begin{center}
    \includegraphics[width=\linewidth]{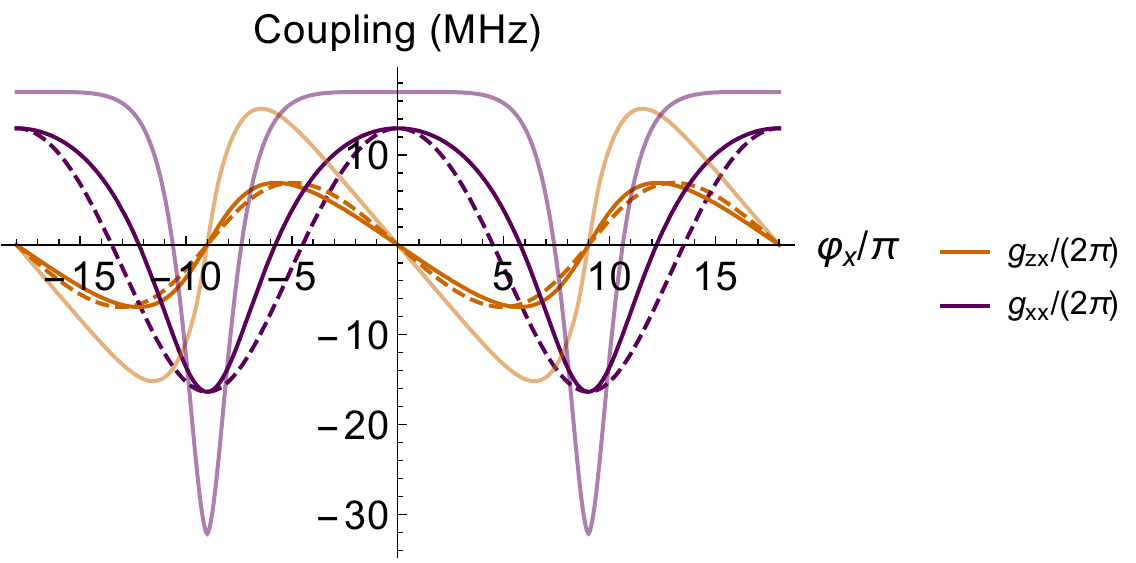}
    \captionof{subfigure}{Longitudinal ($g_{zx}$) and transverse coupling ($g_{xx}$).}  
    \label{fig:coupkn_1}
    \includegraphics[width=\linewidth]{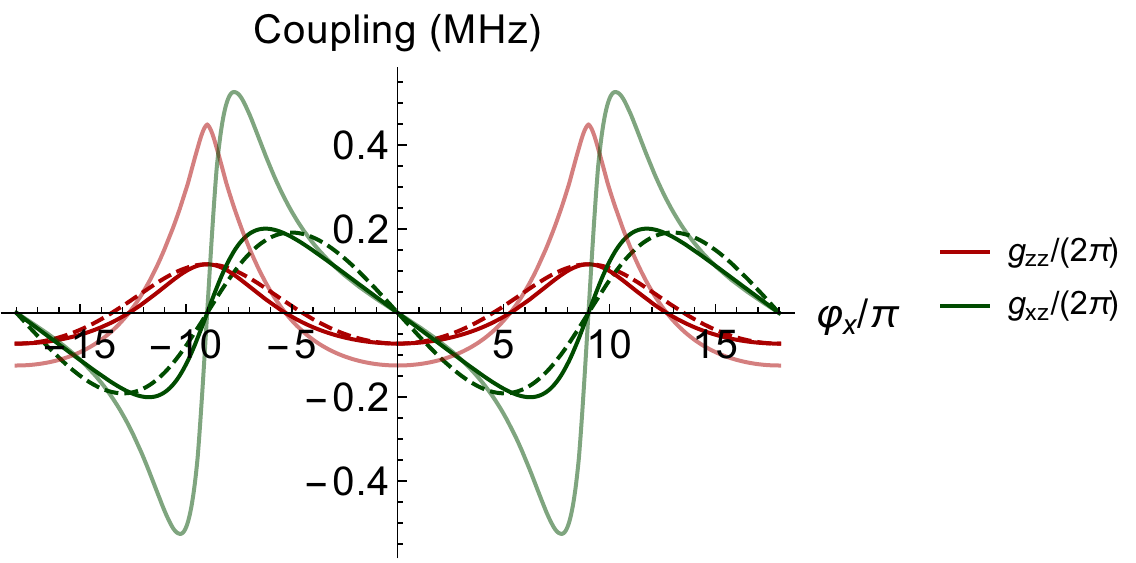}
    \captionof{subfigure}{Unwanted coupling terms, note the change of scale.}  
    \label{fig:coupkn_2}
    \captionof{figure}{The four most important coupling terms as a function of the (reduced) flux through the coupling loops $\varphi_x$ for a coupling junction array of $k = 9$ junctions per array. Solid lines show accurate numerical results, dashed lines show the predictions using the formulas from Sec.~\ref{sec:quant}, the lighter color curves show what happens at a flux $\varphi_{Xb} = \pi$ in the big loop. The longitudinal coupling is almost doubled due to the large-loop flux-biasing, while the point where the transverse coupling disappears is considerably shifted. The longitudinal coupling always disappears exactly at multiples of $\varphi_x = k\, \pi$.}
    \label{fig:coupkn}
    \end{center}
\end{Figure}

With such a treatment, we are of course neglecting the dynamics of the internal degrees of freedom of the array \cite{viola, Hutter2011}. This is justified as long as the energies of these degrees of freedom are far enough separated from the relevant energies of our system, that is the frequencies of qubit and resonator. Explicitly, we have to require their plasma frequencies $\sqrt{8 E_{Ci} E_{Ji}}/h$ to be above 20 GHz in order to push the self-resonant modes of the array well above the resonator mode. Apart from that, we require that $E_{Ji}/E_{Ci} \geq$ 100 to prevent phase slips \cite{Matveev2002, Pop2010}, where $E_{Ci}$ is the charging energy of each individual junction. Putting these two constraints together, we can conclude that each coupling junction in such an array needs to have a Josephson energy larger than $E_{Ji} = h $ 70 GHz, which is a lot bigger than what we assumed for the single-junction case. Apart from this restriction, we proceed just as in the previous section, trying out different parameter values, now including the number of junctions $k$ per coupling array, until finding the \textit{optimal} solution.\\
Table~\ref{tb:kn} shows the chosen parameters for the multi-junction case, here for $k = 9$ junctions per array. While we ascertain that the unwanted coupling terms are considerably suppressed compared to the longitudinal and the transverse coupling, the coupling is smaller in general (see Fig.~\ref{fig:coupkn}). The longitudinal coupling is suppressed by almost one order of magnitude compared to the single-junction case. While the unwanted $g_{xz}$ coupling reaches approximately 4~\% of the longitudinal coupling $g_{zx}$ at their joint maximum, the unwanted $g_{zz}$ coupling reaches only 0.4~\% of the transverse coupling $g_{xx}$ at their maximum.

\begin{Table}
\begin{center}
    \begin{tabular}{| >{\centering}m{.16\linewidth} | c || >{\centering}m{.16\linewidth} | c |}
    \hline
    \multicolumn{2}{|c||}{Parameters} & \multicolumn{2}{c|}{Results}\\
    \hline
    $E_{Jq}$ & $h$ 10 GHz & $\omega_r/(2\pi)$ & 6 - 8 GHz \\ \hline
    $E_{J\Sigma}$ & $h$ 160 GHz & $\Delta/(2\pi)$ & 5.3 - 6.3 GHz\\ \hline
    $E_{J\Delta}/E_{J\Sigma}$ & 0.02 & $g_{zx}^\text{max}/(2\pi)$ & 6 MHz \\ \hline
         $C$ & 102 fF & $g_{xx}^\text{max}/(2\pi)$ & 13 MHz \\ \hline
     $C_q$ & 60 fF & $g_{zz}^\text{max}/(2\pi)$ & 0.07 MHz \\ \hline
   $L$ & 5.0 nH & $g_{xz}^\text{max}/(2\pi)$ &  0.2 MHz \\ \hline
   $L_\text{max}$ & 5.0 nH & $|\alpha_r^{(q)}|$ & 0.9 - 1.5\% \\ \hline
   $L_\text{crit}$ & 5.6 nH & $|\alpha_r^{(r)}|$ &  $\leq$ 0.007\% \\ \hline
    \end{tabular}
    \captionof{table}{The chosen parameters for the case of coupling junction arrays, here for $k=9$, at zero flux through the big loop $\varphi_{Xb} = 0$. $L$ needs to be less or equal to $L_\text{max}$ to ensure that the resonator frequency stays in the 6 - 8 GHz range and less than $L_\text{crit}$ in order to avoid a double-well potential for all possible values of flux.
    On the right the frequencies, anharmonicities, and couplings, which vary with the flux in the coupling loops.}
    \label{tb:kn}
\end{center}
\end{Table}

\begin{Figure}
\def\big{\includegraphics[width=.8\linewidth]{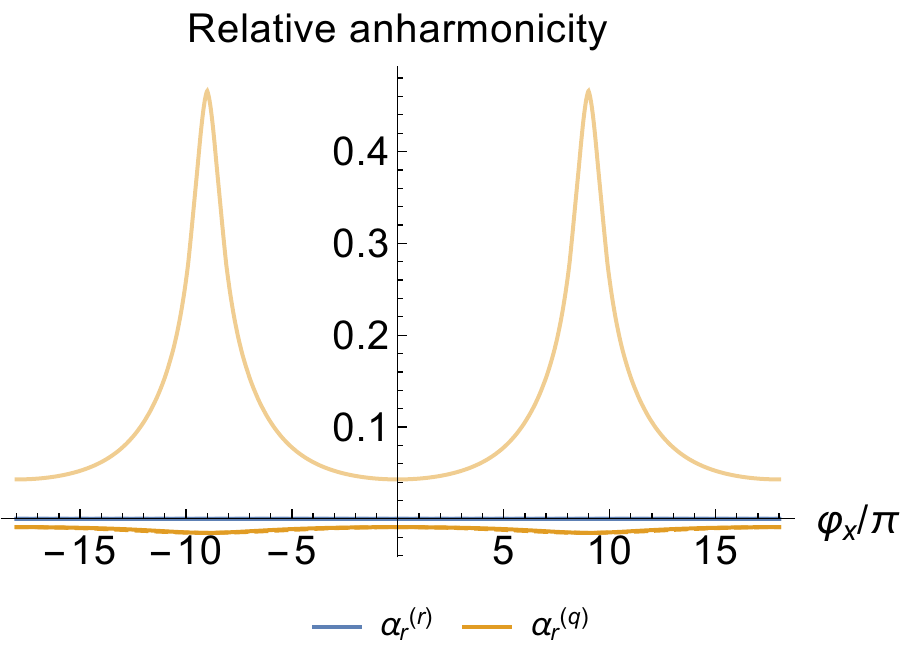}}
\def\little{\includegraphics[width=.38\linewidth]{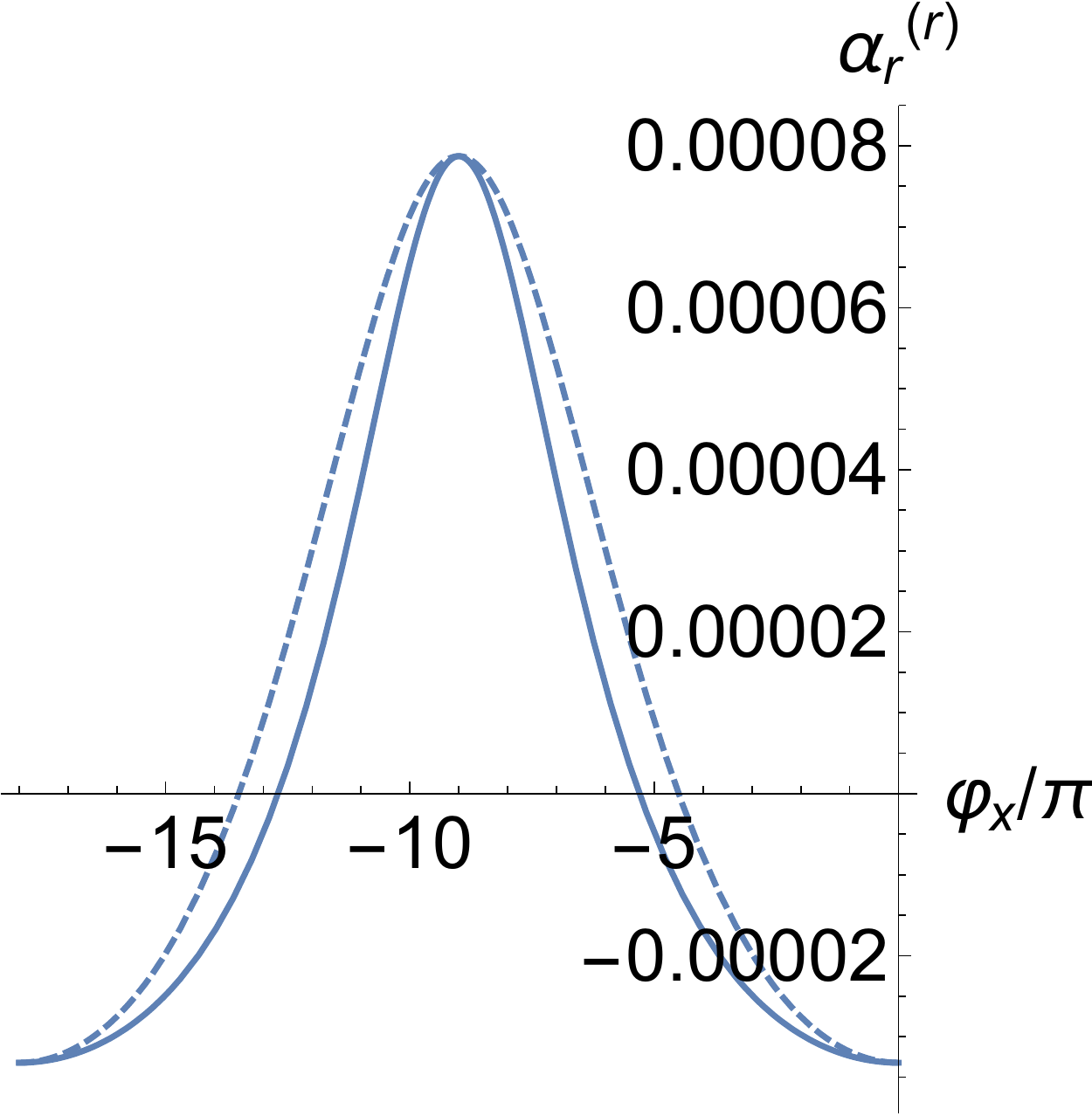}}
\def\stackalignment{r}
\topinset{\little}{\big}{0pt}{-45pt}
    \captionof{figure}{The relative anharmonicities of qubit and resonator as a function of the (reduced) flux through the coupling loops $\varphi_x$ for a coupling junction array with $k = 9$ junctions per array. Solid lines show accurate numerical results, dashed lines show the predictions using the formulas from Sec.~\ref{sec:quant}. Note that in this plot the predictions are indistinguishable from the numerical results. The lighter color curve shows results at a flux $\varphi_{Xb} = \pi$ in the big loop. The smaller plot on the right shows again the relative anharmonicity of the resonator, note the change of scale.}  
    \label{fig:anhkn}
\end{Figure}

The resonator anharmonicity is considerably suppressed to less than 0.007~\%, while the qubit anharmonicity stays roughly the same (see Fig.~\ref{fig:anhkn}). Putting a flux of $\varphi_{Xb} = \pi$ through the big loop has a similar effect as before. The qubit anharmonicity changes sign and is boosted to up to more than 30~\%, while the resonator anharmonicity is not affected by the large-loop flux-biasing.
Even though the suppression of the unwanted coupling and the resonator anharmonicity are a considerable improvement over the single-junction case, the simultaneous suppression of the longitudinal coupling is unfortunate. We will therefore try out another adaptation of the original circuit, as described below.

\section{CASE THREE: CIRCUIT WITH ADDITIONAL INDUCTANCE}
\label{sec:adaptation}

Figure~\ref{fig:add} shows an adaptation of the original circuit in which we have added an additional inductance in each coupling branch, in series with both the coupling junction array and the already existing inductance. 

\begin{Figure}
  \begin{center}
    \includegraphics[width=.95\linewidth]{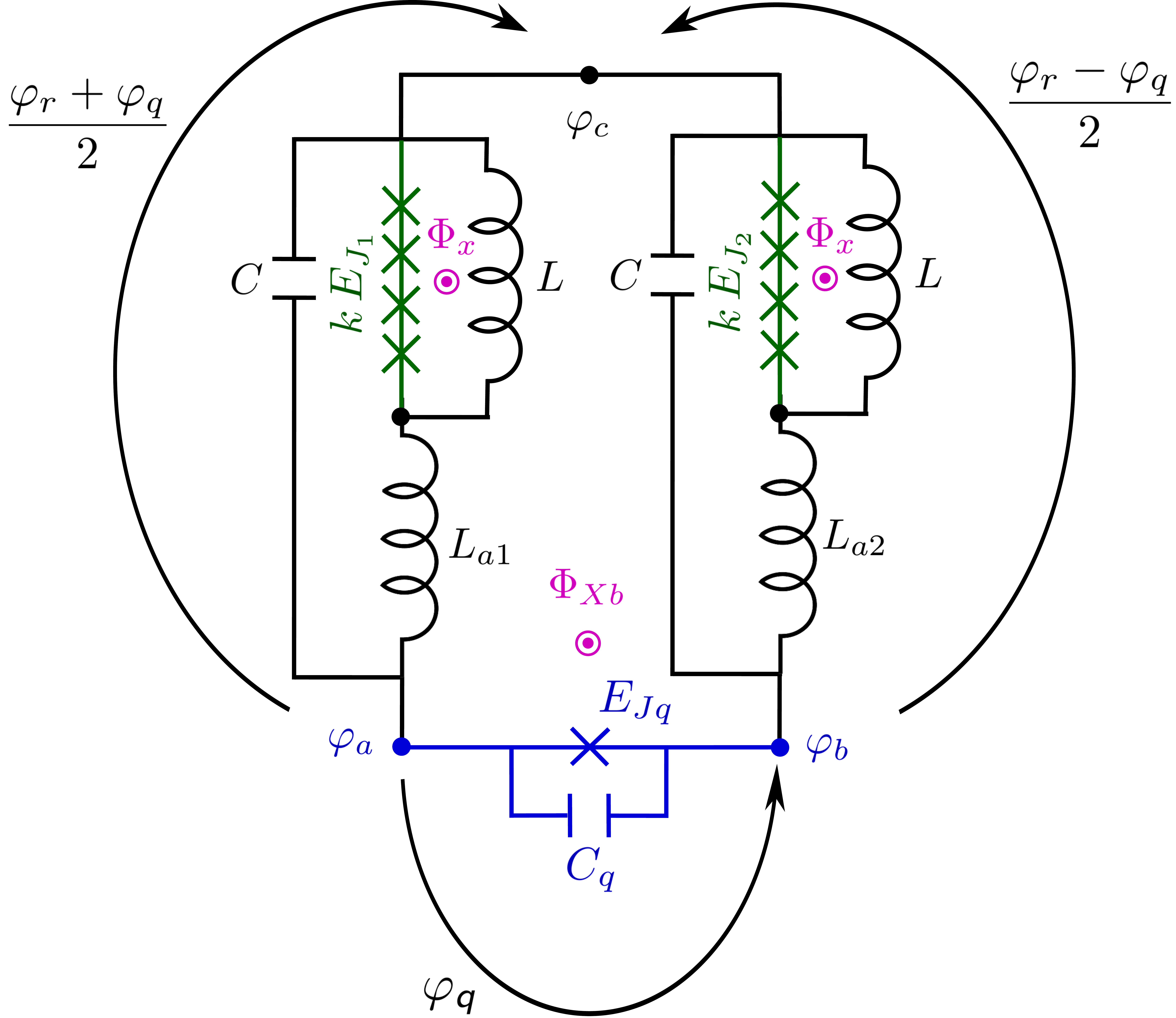}
    \captionof{figure}{Adapted qubit-resonator system with additional inductances $L_{ai}$ in the coupling branches.}  
    \label{fig:add}
    \end{center}
\end{Figure}

Clearly, this adds an additional degree of freedom to each coupling branch. However, this additional variable can be considered to be a dependent variable (just like the internal degrees of freedom within the coupling array) that can be eliminated, as it does not have a significant capacitive term and therefore no low-frequency dynamics on its own. 

\begin{wrapfigure}{r}{0.6\linewidth}
\vspace{-20pt}
  \begin{center}
    \includegraphics[width=.9\linewidth]{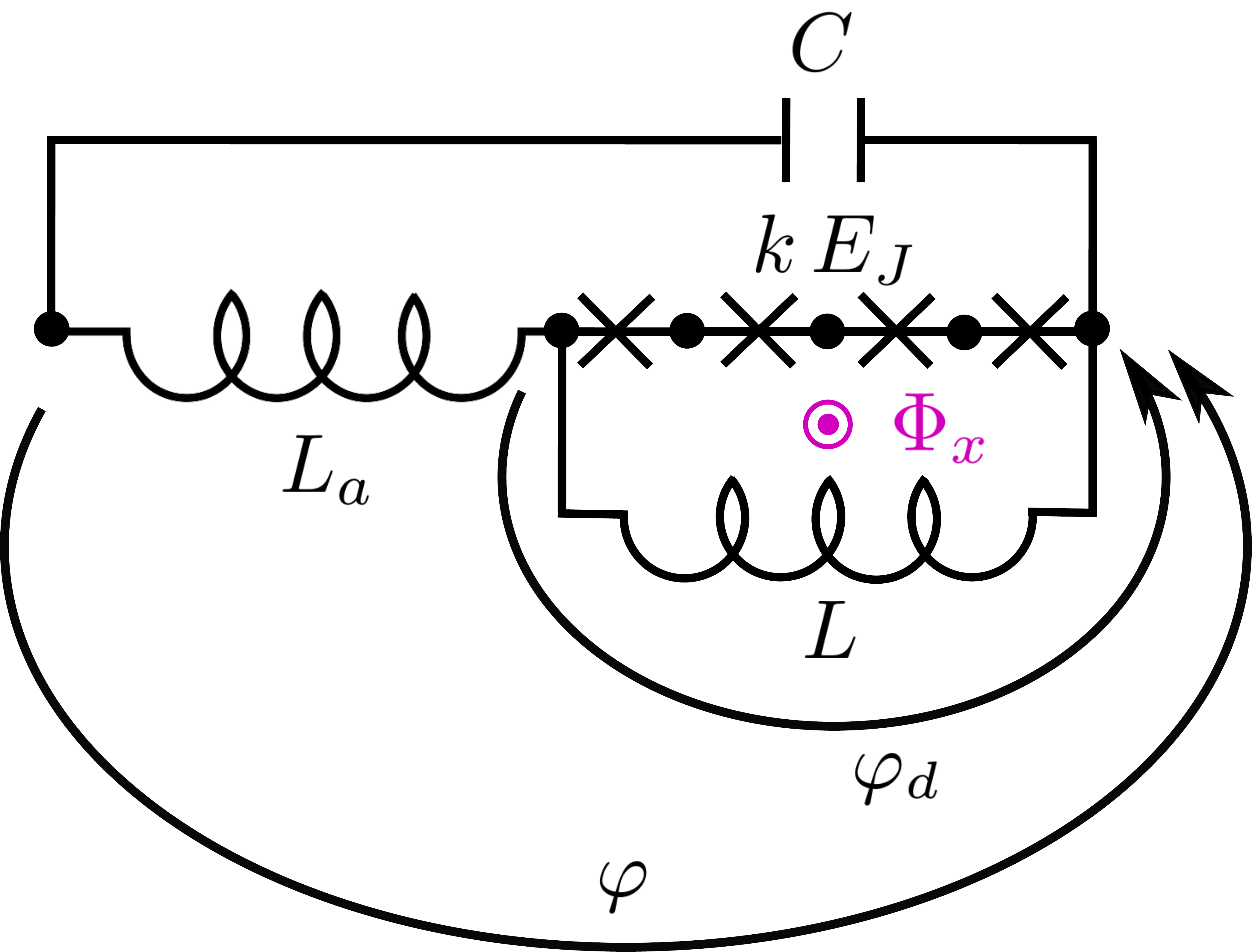}
  \end{center}
  \vspace{-10pt}
   \caption{Detail of a single coupling branch from Fig.~\ref{fig:add}. The phase difference across the junction array $\varphi_d$ has no dynamics on its own but depends on the phase difference $\varphi$ across the whole device.}
    \label{fig:junc_ind}
\end{wrapfigure}

To explain this, we will start with a description of a single coupling branch as shown in Fig.~\ref{fig:junc_ind}. 
This is a system with $n=k+2$ nodes, where $k$ is the number of junctions in the array. That makes $n-1=k+1$ degrees of freedom, $k$ of them without their own dynamics. We define $\varphi$ as the phase difference across the whole device and denote the dependent phase difference over the junction array as $\varphi_d$ as depicted in Fig.~\ref{fig:junc_ind}, the phase difference over a single junction being $\varphi_d/k$, assuming the junctions are all equal and $m = 0$ in Eq.~\ref{eq:u_array}. This already eliminates all phases inside the junction array. The phase difference across the inductance $L_a$ must then be $\varphi-\varphi_d$. The Lagrangian for the system shown in Fig.~\ref{fig:junc_ind} yields

\begin{align*}
\mathcal{L} = \left(\frac{\Phi_0}{2\pi}\right)^2 &\left(\frac{C}{2} \dot \varphi^2 - \frac{1}{2L_a} (\varphi - \varphi_d)^2 - \frac{1}{2L} \varphi_d^2\right)\\
& + k \, E_J \cos\left(\frac{\varphi_d - \varphi_x}{k}\right), \numberthis
\end{align*}

where $\varphi_x = 2\pi \, \Phi_x/\Phi_0$ is again the reduced external flux through the coupling loop. From here we can deduce the equations of motion for $\varphi$ and $\varphi_d$, being

\begin{align*}
C \ddot \varphi &=  \frac{1}{L_a} (\varphi_d - \varphi) \numberthis \\
0 &= \frac{1}{L_a} (\varphi_d - \varphi) + \frac{1}{L} \varphi_d + \left(\frac{2\pi}{\Phi_0}\right)^2 E_J \sin\left(\frac{\varphi_d - \varphi_x}{k}\right).
\end{align*}

As noted above, there is no capacitive term in the second equation. It is thus not a differential equation, but simply a non-linear algebraic equation in $\varphi$ and $\varphi_d$, which can be used to eliminate $\varphi_d$. However, it is not analytically possible to solve the second equation for $\varphi_d$. Our strategy will therefore be to solve it for $\varphi$ and invert this function numerically for a given set of parameters in order to eliminate $\varphi_d$. Solving for $\varphi$ thus yields

\begin{align}
\varphi(\varphi_d) = \gamma \, \varphi_d + \beta \sin\left(\frac{\varphi_d - \varphi_x}{k}\right)
\label{eq:invert}
\end{align}

with the abbreviations $\beta = (2\pi/\Phi_0)^2 L_a E_J$ and $\gamma = 1 + L_a/L$, where $\beta$ corresponds to the screening parameter known from SQUID terminology \cite{clarke}, that is the ratio between Josephson energy and inductive energy. 

\begin{Figure}
  \begin{center}
    \includegraphics[width=\linewidth]{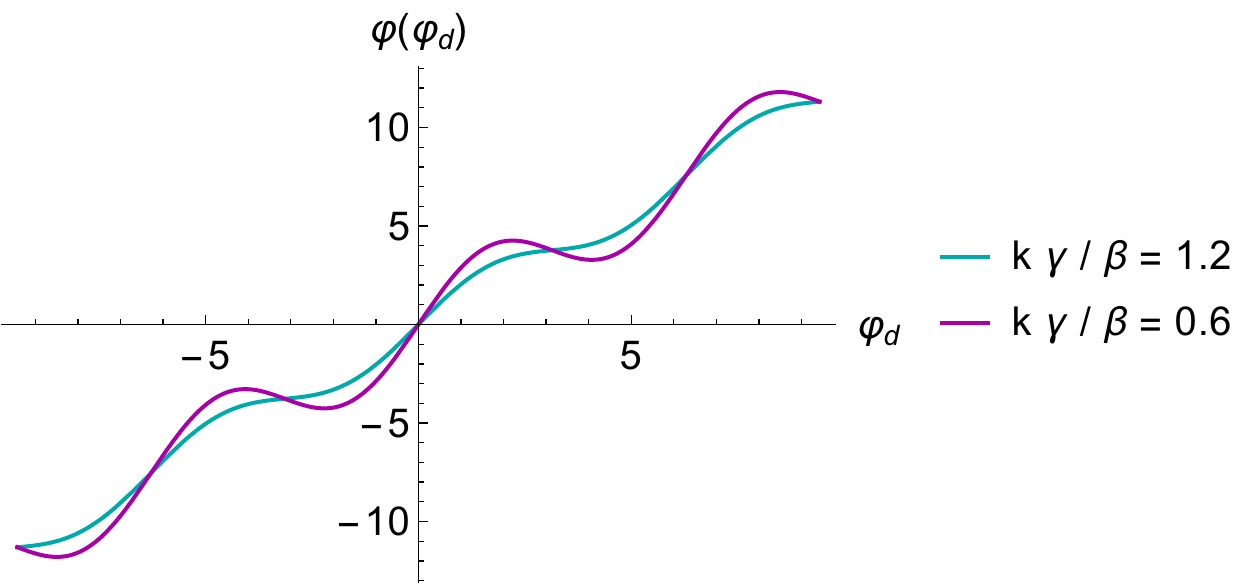}
    \captionof{figure}{$\varphi$ as a function of the dependent variable $\varphi_d$ (referring to Fig.~\ref{fig:junc_ind}) for different parameters. While the blue curve with $k\, \gamma/\beta > 1$ is invertible, the magenta one with $k\, \gamma/\beta < 1$ is not.}  
    \label{fig:invertible}
    \end{center}
\end{Figure}

In terms of the parameters $k$, $\beta$ and $\gamma$, we can distinguish two different cases. The function is invertible as long as $k\, \gamma/\beta$ is above the critical value of one, compare Fig.~\ref{fig:invertible}. If the function is not invertible, the potential becomes multi-valued. This is a parameter regime we want to avoid. In order to see what that condition means, we can rewrite it as

\begin{align}
\gamma > \beta/k \quad \Leftrightarrow \quad E_{L} + E_{La} > E_J/k,
\label{eq:crit}
\end{align}

where $E_L = (\Phi_0/(2\pi))^2/L$ is the energy associated with the inductance and $E_{La}$ is the same for the additional inductance $L_a$. The condition given in Eq.~\ref{eq:crit} thus means that the energy of the two inductances must be bigger than the energy of the junction array, in order to ensure that Eq.~\ref{eq:invert} is invertible and there is a well-defined potential.\\
The potential energy for the complete qubit-resonator system depicted in Fig.~\ref{fig:add} is given by

\begin{align*}
\mathcal{U} &= \left(\frac{\Phi_0}{2\pi}\right)^2 \biggl(\frac{1}{2L_{a1}} f\left(\frac{\varphi_r + \varphi_q}{2}, \varphi_x, \beta_1, k, \gamma_1\right)  \\
&+\frac{1}{2L_{a2}} f\left(\frac{\varphi_r - \varphi_q}{2}, \varphi_x, \beta_2, k, \gamma_2\right)\biggr) \\
&- E_{Jq} \cos(\varphi_q + \varphi_{Xb}), \numberthis
\end{align*}

where

\begin{align*}
f(\varphi, \varphi_x, \beta, k, \gamma) = \varphi^2 &- 2\, \varphi\, \varphi_d + \gamma \,\varphi_d^2 \\
&- 2 \,k \,\beta \cos\left(\frac{\varphi_d - \varphi_x}{k}\right) \numberthis
\end{align*}

is a function that describes one coupling branch as depicted in Fig.~\ref{fig:junc_ind}, in which the dependent variable $\varphi_d$ must be replaced by the numerical inversion of Eq.~\ref{eq:invert}. The kinetic energy is the same as for the original circuit (Eq.~\ref{eq:kin}). In analogy to what we described in Sec.~\ref{sec:fluxbias}, we want this circuit to be also usable at a flux-biasing of $\varphi_{Xb} = \pi$ through the big loop. We thus have to make sure that we do not go into parameter ranges, where the potential is a double well. This can be done by determining the curvature of the potential in the $\varphi_q$ direction at a flux $\varphi_{Xb} = \pi$ through the big loop and $\varphi_x = k\, \pi$ through the coupling loops. If the curvature is positive here, it will always be positive. While we can not define a critical inductance as done in Sec.~\ref{sec:fluxbias} (Eq.~\ref{eq:lcrit}), the equivalent in this case is a critical (minimum) number of junctions $k_\text{crit}$ that ensures a positive curvature of the potential.\\
From here on, our strategy is the one described in Sec.~\ref{sec:parameters}. For a given set of parameters (including the external fluxes), we first determine the position of the potential energy minimum in $\varphi_q$ and $\varphi_r$ and then calculate the frequencies, anharmonicities, and couplings using series approximations around that minimum. To choose the best parameters, we again fix the capacitance $C$ in terms of the other variables, such that the resonator frequency is $\omega_r/(2\pi) = 8$ GHz at zero flux, where $\omega_r$ has its maximum. Then we try out different values for the other parameters until we find the solution that gives the highest longitudinal coupling and anharmonicity, while satisfying all the conditions mentioned above. The chosen parameters for this circuit are shown in Tab.~\ref{tb:add}.

\begin{Table}
\begin{center}
    \begin{tabular}{| >{\centering}m{.16\linewidth} | c || >{\centering}m{.16\linewidth} | c |}
    \hline
    \multicolumn{2}{|c||}{Parameters} & \multicolumn{2}{c|}{Results}\\
    \hline
    $E_{Jq}$ & $h$ 5 GHz & $\omega_r/(2\pi)$ & 6 - 8 GHz \\ \hline
    $E_{J\Sigma}$ & $h$ 155 GHz & $\Delta/(2\pi)$ & 4.8 - 5.8 GHz\\ \hline
    $E_{J\Delta}/E_{J\Sigma}$ & 0.02 & $g_{zx}^\text{max}/(2\pi)$ & 10 MHz \\ \hline
    $C$ & 65 fF & $g_{xx}^\text{max}/(2\pi)$ & 9 MHz \\ \hline
    $C_q$ & 50 fF & $g_{zz}^\text{max}/(2\pi)$ & 0.06 MHz \\ \hline
    $L$ & 4.5 nH & $g_{xz}^\text{max}/(2\pi)$ & 0.5 MHz \\ \hline
    $L_a$ & 3 nH &  $|\alpha_r^{(q)}|$ & 1.1 - 2\%  \\ \hline
     $k_\text{crit}$ & 3.3 & $|\alpha_r^{(r)}|$ &  $\leq$ 0.003\%  \\ \hline
    \end{tabular}
    \captionof{table}{The chosen parameters for the case of the adapted circuit with the added inductance, here for $k=5$ junctions per coupling array, at zero flux through the big loop $\varphi_{Xb} = 0$. $k_\text{crit}$ defines a lower threshold for the number of junctions $k$ in order to avoid a double-well potential for all possible values of flux.
    On the right the frequencies, anharmonicities, and couplings, which vary with the flux in the coupling loops.}
    \label{tb:add}
\end{center}
\end{Table}

Figure~\ref{fig:addfreq} shows the frequencies of qubit and resonator as a function of the flux through the coupling loops $\varphi_x$. Their flux dependence looks a lot like in the two cases described above. The lighter color curve shows results at a flux $\varphi_{Xb} = \pi$ in the big loop. The qubit frequency again experiences a drop due to the large-loop flux-biasing, while the resonator is unaffected by this.

\begin{Figure}
  \begin{center}
    \includegraphics[width=\linewidth]{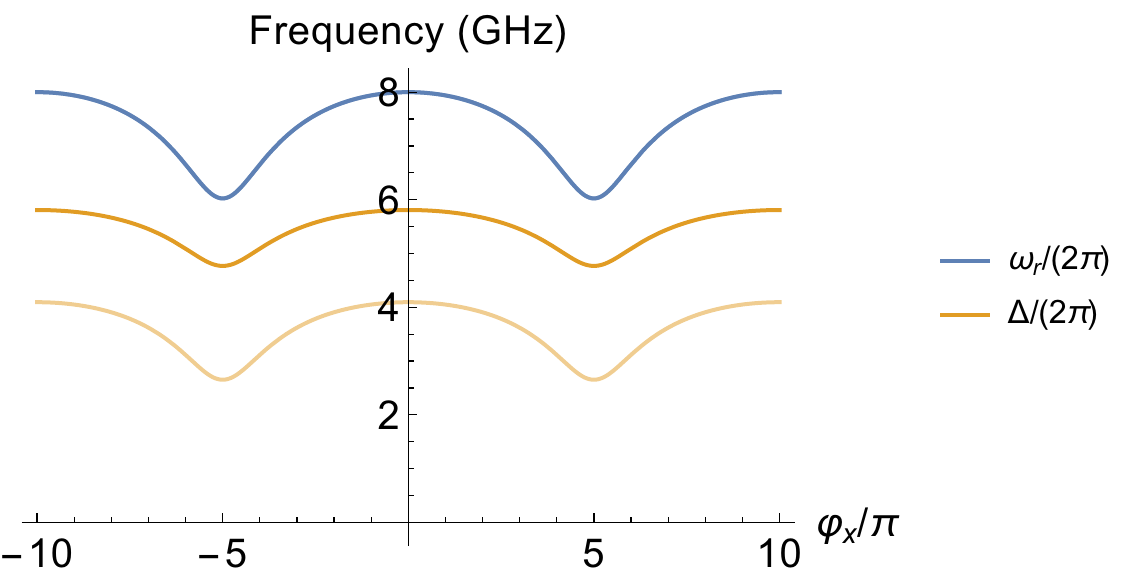}
    \captionof{figure}{The frequencies of qubit and resonator as a function of the (reduced) flux through the coupling junctions $\varphi_x$ for the adapted circuit with $k = 5$ junctions per array. The lighter color curve shows results at a flux $\varphi_{Xb} = \pi$ in the big loop.}  
    \label{fig:addfreq}
    \end{center}
\end{Figure}

Figure~\ref{fig:addcoup} shows the four most important coupling terms, again as a function of the flux through the coupling loops $\varphi_x$. Compared to the single-junction case, the coupling is smaller and loses its resemblance to the trigonometric functions in the formulas given in Sec.~\ref{sec:quant}. The unwanted coupling terms are suppressed. The unwanted $g_{xz}$ coupling reaches 4~\% of the longitudinal coupling at their joint maximum, while the unwanted $g_{zz}$ coupling is about 0.5~\% of the transverse coupling at zero flux. Though the longitudinal coupling $g_{zx}$ is smaller than in the single-junction case, it is slightly bigger than in the case of the coupling junction array without the additional inductance. The lighter color curves show what happens at a flux $\varphi_{Xb} = \pi$ in the big loop. Due to the large-loop flux-biasing, the longitudinal coupling is almost doubled. The point where the transverse coupling disappears is considerably shifted, along with the maximum of the longitudinal coupling. 

\begin{Figure}
  \begin{center}
    \includegraphics[width=\linewidth]{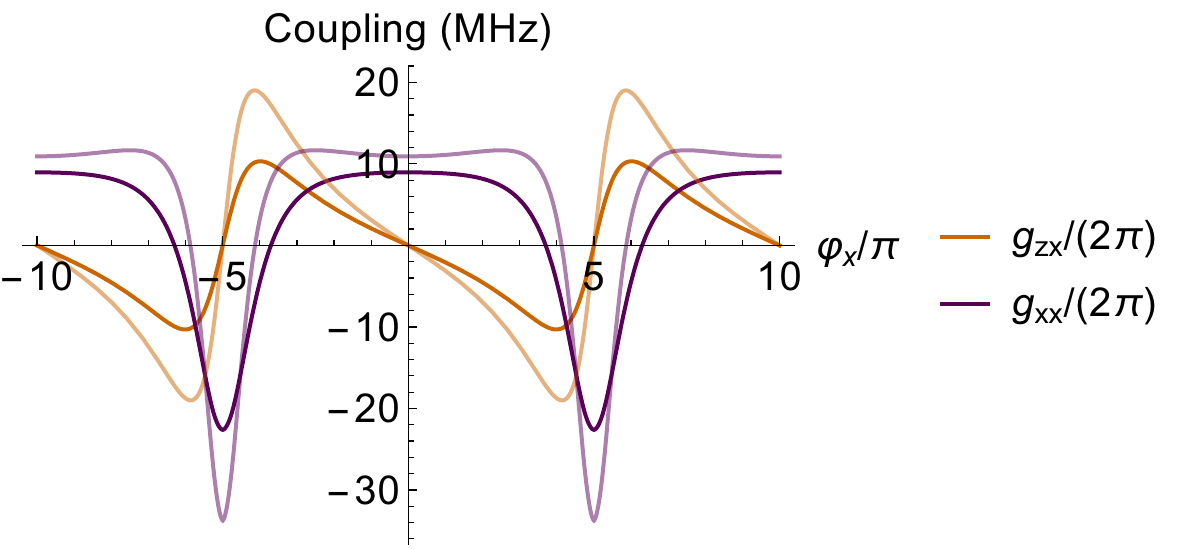}
    \captionof{subfigure}{Longitudinal ($g_{zx}$) and transverse coupling ($g_{xx}$).}  
    \label{fig:addcoup_1}
    \includegraphics[width=\linewidth]{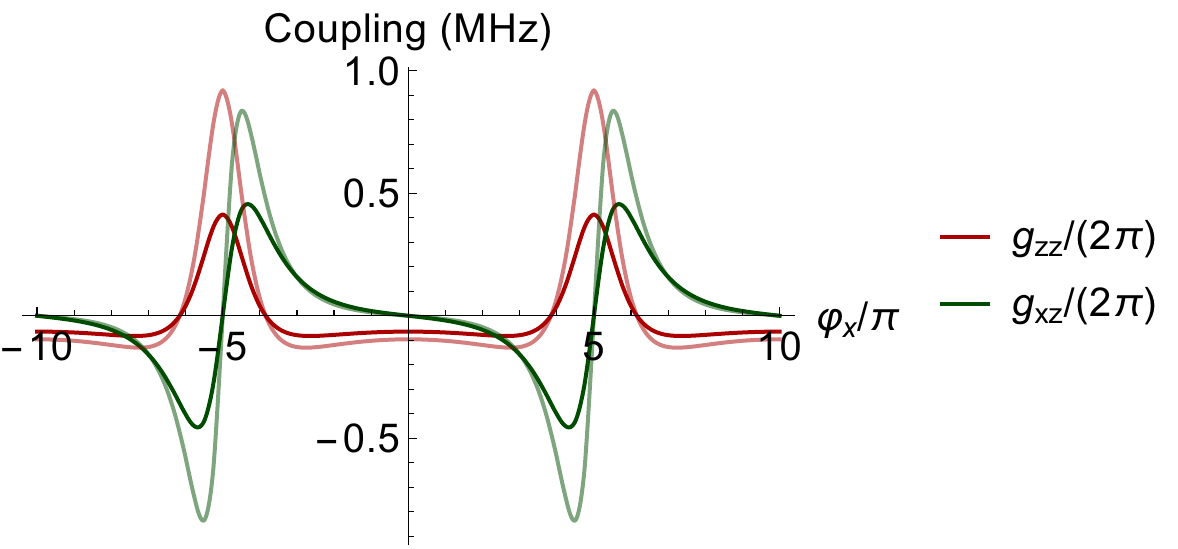}
    \captionof{subfigure}{Unwanted coupling terms, note the change of scale.}  
    \label{fig:addcoup_2}
    \captionof{figure}{The four most important coupling terms as a function of the (reduced) flux through the coupling junctions $\varphi_x$ for the adapted circuit with $k = 5$ junctions per array. The lighter color curves show what happens at a flux $\varphi_{Xb} = \pi$ in the big loop. Due to the large-loop flux-biasing, all coupling terms get slightly bigger (the longitudinal coupling is almost doubled), while the point where the transverse coupling disappears is considerably shifted.}
    \label{fig:addcoup}
    \end{center}
\end{Figure}

\begin{Figure}
\def\big{\includegraphics[width=.8\linewidth]{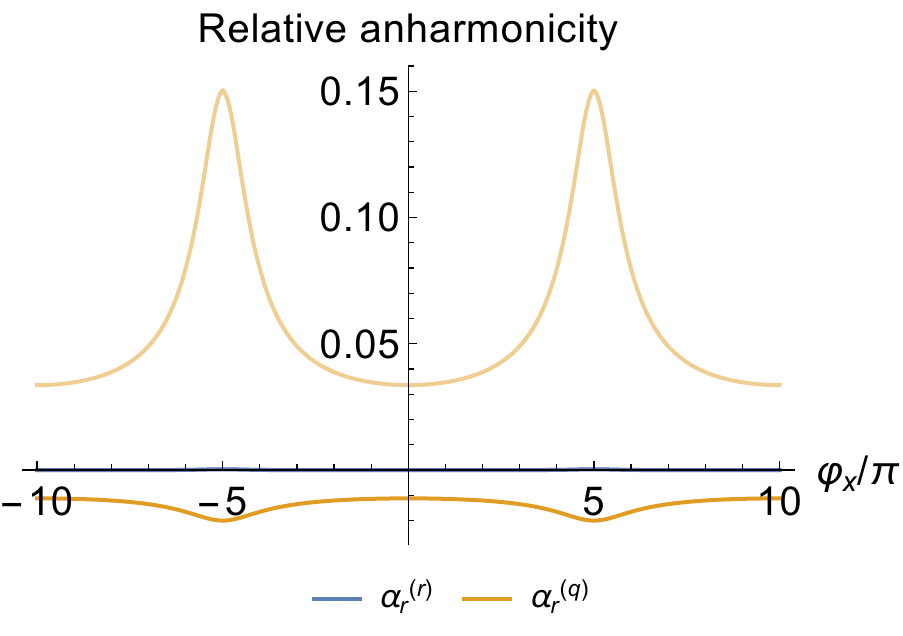}}
\def\little{\includegraphics[width=.38\linewidth]{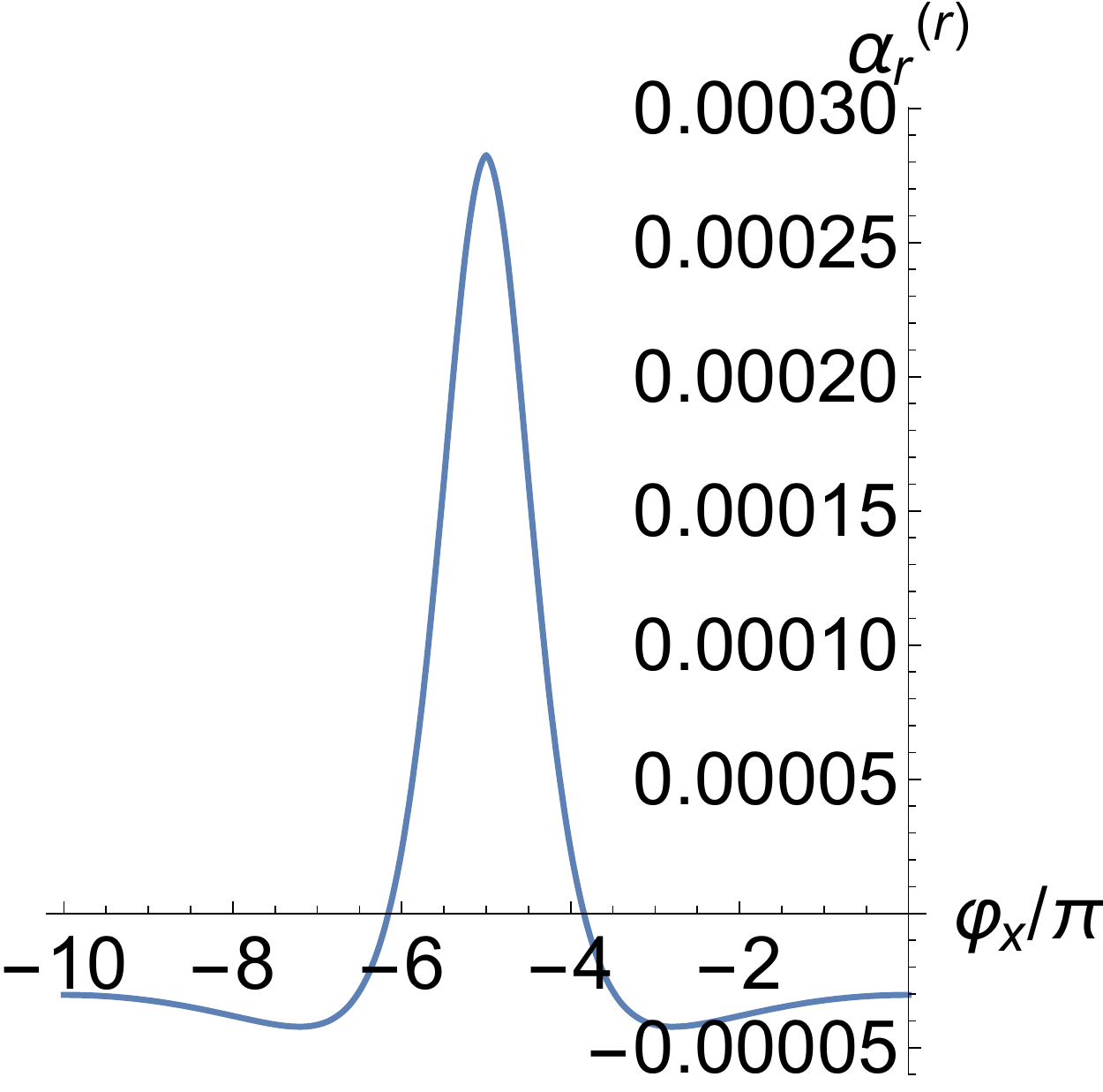}}
\def\stackalignment{r}
\topinset{\little}{\big}{-10pt}{-45pt}
    \captionof{figure}{The relative anharmonicities of qubit and resonator as a function of the (reduced) flux through the coupling loops $\varphi_x$ for the adapted circuit with $k = 5$ junctions per array. The lighter color curve shows results at a flux $\varphi_{Xb} = \pi$ in the big loop. While the resonator anharmonicity is unchanged, the qubit anharmonicity is now positive and boosted to up to 18~\%. The smaller plot on the right shows again the relative anharmonicity of the resonator, note the change of scale.}  
    \label{fig:addanh}
\end{Figure}

The qubit anharmonicity is slightly bigger than in the single-junction or the coupling array case, while the resonator anharmonicity is suppressed to less than 0.03~\% (see Fig.~\ref{fig:addanh}). The large-loop flux-biasing leads again to a boost in qubit anharmonicity, here to up to 18~\%. 
We can conclude that the adapted circuit with the additional inductance works better than the circuit using only the junction array. The single-junction case seems problematic due to the high resonator anharmonicity. In all cases flux-biasing with $\varphi_{Xb} = \pi$ in the big loop leads to a boost in anharmonicity and to an increase of coupling strength of almost a factor of two.

\section{PHYSICAL IMPLEMENTATION}
\label{sec:implementation}

\begin{figure*}[ht]
	\centering
	\includegraphics[width=\textwidth]{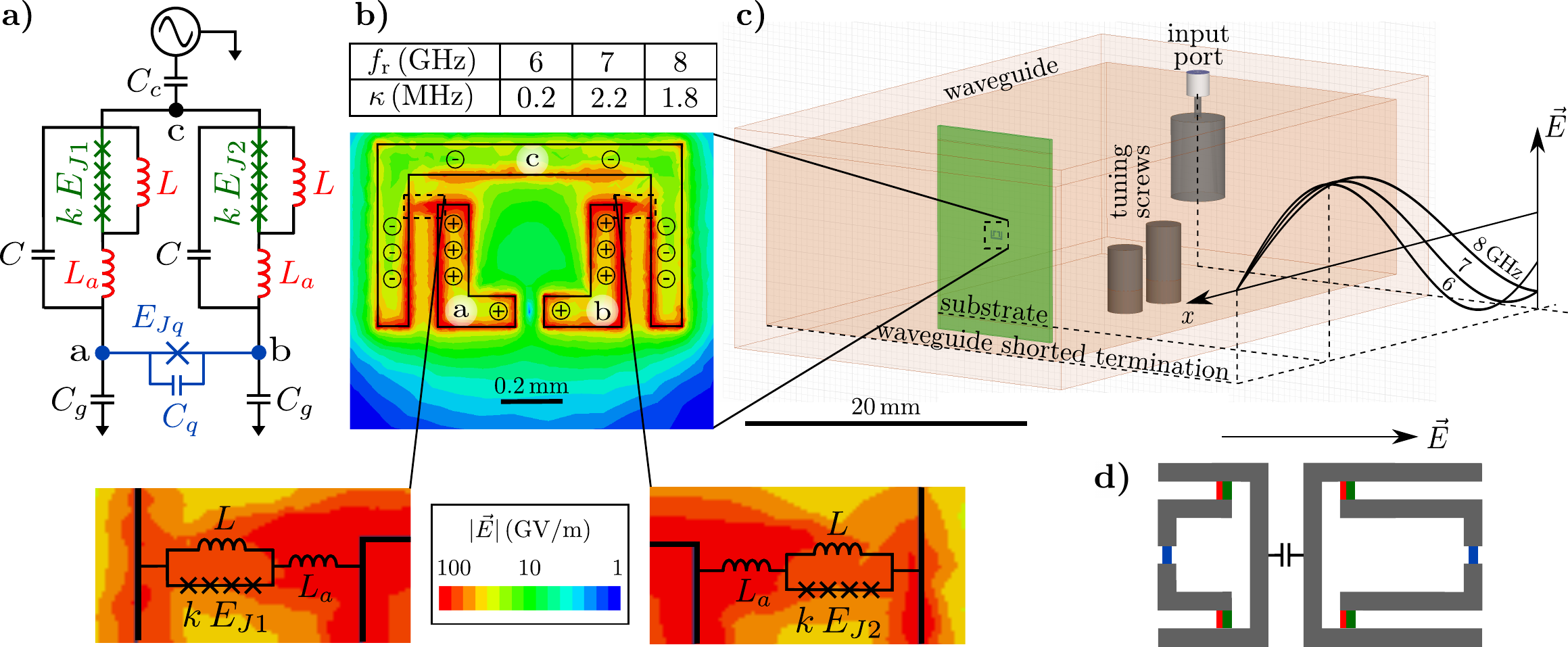}
	\caption{Proposal for the physical implementation of the inductively shunted transmon qubit with tunable transverse and longitudinal coupling. \textbf{a)} Electrical schematic of the qubit-resonator circuit coupled to the control and readout microwave environment. The qubit dynamics is dominated by the Josephson junction with energy $E_{Jq}$ and capacitance $C_q$ (colored in blue). The frequency of the resonator mode is given by the equivalent inductance formed by $L$, $L_a$ (colored in red), the inductances of the Josephson junction arrays (colored in green), and the shunting capacitances $C$. \textbf{b)} Finite element model used to simulate the resonator coupling to the microwave drives. The color scale indicates the magnitude of the computed electric field at the surface of the thin film superconducting electrodes, for a total energy stored in resonator mode of $1\,\mathrm{J}$. The $\mathbf{+}$ and $\mathbf{-}$ symbols represent the polarity of the electric field. The capacitors $C$ and $C_q$ are implemented using so-called finger capacitors, while $C_c$ and $C_g$ are given by the stray field coupling to the rectangular waveguide sample holder shown in Panel \textbf{c}. The inductive elements of the circuit are introduced in the model as lumped elements connecting the pads (shown in the insets below). The table shows the resulting linewidth values $\kappa$ for three different frequencies of the resonator mode, chosen in the pass-band of the waveguide. \textbf{c)} Finite element model used to simulate the 3D waveguide sample holder. Recently, a similar sample holder geometry has been used to perform multiplexed quantum readout \cite{Kou2017}. The qubit-resonator circuit is deposited on a sapphire substrate which is indicated by the green rectangle. The electric field magnitude along the waveguide is frequency dependent, and its profile is schematically shown for $6$, $7$, and $8\,\mathrm{GHz}$. The impedance and the mode profile between the waveguide and the coaxial cable connected to the input port are matched using the tuning screws. \textbf{d)} Direct extension of the proposed physical implementation for two capacitively coupled qubit-resonator systems (compare Ref.~\cite{richer}). The resonators are designed to have different eigenmode frequencies, and they can be individually addressed using the collective waveguide mode represented by the direction of the $\vec{E}$ field.}
\label{fig:phys_impl}
\end{figure*}

Figure~\ref{fig:phys_impl} shows a possible physical implementation of the inductively shunted transmon qubit. One of the main challenges is to realize compact, low-loss and linear inductances, in the range of several $\mathrm{nH}$, required for the shunting inductors $L$ and $L_a$ (see Fig.~\ref{fig:phys_impl}\textbf{a}). For this purpose, we propose the use of a superconducting strip consisting of a high kinetic inductance material such as granular aluminum, or niobium and titanium nitrides, which have been shown to achieve inductances in the range of $\mathrm{nH}/\square$ \cite{Rotzinger2017, Annunziata2010, Samkharadze2016, Vissers2010}. The rest of the circuit, including all Josephson junctions, can be fabricated using standard thin-film aluminum. The electrical connections between these different metallic layers can be realized using recently developed argon ion cleaning and contacting techniques which preserve the coherence of the circuit \cite{Dunsworth2017, Wu2017, Grunhaupt2017}. \\
The capacitances required for shunting the qubit, $C_q$, and the resonator, $C$, as well as the coupling capacitors, $C_c$ and $C_g$, can all be implemented by the relatively simple structure shown in Fig.~\ref{fig:phys_impl}\textbf{b}. For clarity, the three superconducting island phases are labeled using the same notation as in Fig.~\ref{fig:qubit_resonator}. The structure is designed to couple to the first propagating mode of a 3D wave\-guide, following the sample-holder geometry described in Ref.~\cite{Kou2017}.  The electric-field magnitude is indicated by the color scale. The maximum values, in the range of $100\,\mathrm{nV/m}$ for an energy of $1\,\mathrm{J}$ stored in the mode, are comparable to the electric field values reported in Ref.~\cite{Grunhaupt2017}, which enabled the measurement of microwave resonators with internal quality factors exceeding $10^6$ in the quantum regime. Notice that the proposed implementation satisfies the required left-right symmetry of the schematics in Fig.~\ref{fig:phys_impl}\textbf{a}, with comfortable margins of error, below $1\,\%$, for either optical or electron-beam lithography.\\
The 3D wave\-guide model shown in Fig.~\ref{fig:phys_impl}\textbf{c} offers the advantage of strong coupling for the resonator mode inside the designed pass band between $6$ and $8\,\mathrm{GHz}$, as indicated by the table in Fig.~\ref{fig:phys_impl}\textbf{b}, while the qubit mode can be efficiently decoupled from the microwave environment. The finite-element simulations indicate a qubit mode coupling quality factor as high as $10^8$.\\
The magnetic field required to tune the fluxes $\Phi_x$ and $\Phi_{Xb}$ (see Fig.~\ref{fig:qubit_resonator}) can be controlled using a direct current coil, which can be attached to the exterior of the sample holder, with the current flowing in a plane perpendicular to the x-axis. In the simplest implementation, the same coil can bias both fluxes, making use of a large ratio between the areas of the superconducting loops enclosing $\Phi_x$ and $\Phi_{Xb}$. Thus, small field variations can be used to tune $\Phi_{Xb}$, quasi-independently from $\Phi_x$.\\
The currently proposed physical implementation is meant as a prototype to test the tunability of the transverse and longitudinal coupling, nevertheless the design shown in Fig.~\ref{fig:phys_impl}\textbf{c} could be adapted for a higher density of qubits. In Fig.~\ref{fig:phys_impl}\textbf{d} we show a direct extension of the concept for two qubits using capacitive coupling between the resonators. With more involved RF designs, it is possible to enlarge the qubit matrix, and add strictly local qubit and resonator drives by using recent advancements in flip-chip and micromachined superconducting circuit technology \cite{Minev2016, Brecht2017, Rosenberg2017}.

\end{multicols}

%
%

\begin{multicols}{2}

\section{SUMMARY}

In conclusion, we presented an inductively shunted transmon qubit design that can be tuned between pure transverse and pure longitudinal coupling to an embedded resonator mode, by changing the external magnetic flux. We performed quantitative analytical and numerical calculations for several qubit-resonator coupling designs. We found that by applying an additional magnetic flux through the loop of the inductively shunted qubit, both the coupling terms and the qubit anharmonicity increase significantly. Additionally, we showed that using single Josephson junctions in the qubit-resonator coupling elements is not feasible, because of the resulting large unwanted coupling terms and high resonator anharmonicity. Using junction arrays in the coupling elements is more favorable, because the ratio between the longitudinal coupling and the unwanted coupling terms can be increased by an order of magnitude, and the resonator anharmonicity is strongly suppressed. Including an additional inductance in the coupling branches helps to further increase the qubit anharmonicity and the longitudinal coupling by up to a factor of two. Finally, we proposed a prototype design based on standard circuit fabrication, integrated with high kinetic inductance elements.

\section*{ACKNOWLEDGMENTS}
SR and DD acknowledge support from the Alexander von Humboldt foundation. DD acknowledges support from Intelligence Advanced Research Projects Activity (IARPA) under Contract No. W911NF-16-0114. 
NM, STS and IMP acknowledge funding from the Alexander von Humboldt foundation in the framework of a Sofja Kovalevskaja award endowed by the German Federal Ministry of Education and Research, and from the Initiative and Networking Fund of the Helmholtz Association, within the Helmholtz Future Project \textit{Scalable solid state quantum computing}.


\vspace{.75cm}

\hrulefill

\vspace{.75cm}


\small{\bibliographystyle{unsrt}
\bibliography{Bibliography}\vspace{0.75in}}

\end{multicols}

\end{document}